\documentclass{aa}  

\usepackage{placeins}
\usepackage{graphicx}
\usepackage{xcolor}
\usepackage{lipsum}
\usepackage{txfonts}
\begin{document}

   \title{LOFAR high-band antenna observations of the Perseus cluster}

   \subtitle{The discovery of a giant radio halo}

  \author{R.~J.~van~Weeren \inst{1}
           \and R.~Timmerman\inst{1,2,3} 
           \and V.~Vaidya\inst{1}
           \and M.-L.~Gendron-Marsolais\inst{4}
           \and A.~Botteon\inst{5}
           \and I.~D.~Roberts\inst{1,6,7}
           \and J.~Hlavacek-Larrondo\inst{8}
           \and A.~Bonafede\inst{9,5}
           \and M.~Br\"uggen \inst{10}
           \and G.~Brunetti \inst{5}
           \and R.~Cassano\inst{5}
           \and V.~Cuciti\inst{9,5}
           \and A.~C.~Edge\inst{2}
             \and F.~Gastaldello \inst{11}
           \and C.~Groeneveld\inst{1}
           \and T.~W.~Shimwell \inst{12,1} 
          }

   \institute{Leiden Observatory, Leiden University, PO Box 9513, 2300 RA Leiden, The Netherlands\\
              \email{rvweeren@strw.leidenuniv.nl}
   \and
   Centre for Extragalactic Astronomy, Department of Physics, Durham University, Durham DH1 3LE, UK
   \and Institute for Computational Cosmology, Department of Physics, Durham University, South Road, Durham DH1 3LE, UK
   \and  D\'epartement de physique, de g\'enie physique et d'optique, Universit\'e Laval, Qu\'ebec (QC), G1V 0A6, Canada 
   \and
   INAF -- Istituto di Radioastronomia, via P. Gobetti 101, 40129 Bologna, Italy
   \and
   Department of Physics \& Astronomy, University of Waterloo, Waterloo, ON N2L 3G1, Canada
   \and
   Waterloo Centre for Astrophysics, University of Waterloo, Waterloo, ON N2L 3G1, Canada 
   \and   
   D\'epartement de Physique, Universit\'e de Montr\'eal, Succ. Centre-Ville, Montr\'eal, Qu\'ebec (QC), H3C 3J7, Canada
   \and
   Dipartimento di Fisica e Astronomia, Universit\`a di Bologna, via P. Gobetti 93/2, I-40129 Bologna, Italy 
   \and
   University  of  Hamburg,  Hamburger  Sternwarte,  Gojenbergsweg 112, 21029 Hamburg, Germany \and   
   INAF-IASF Milano, Via A. Corti 12, I-20133, Milano, Italy
   \and
   ASTRON, The Netherlands Institute for Radio Astronomy, Postbus 2, 7990 AA Dwingeloo, The Netherlands }



\abstract{The Perseus cluster is the brightest X-ray cluster in the sky and is known as a cool-core galaxy cluster. Being a very nearby cluster, it has been extensively studied. This has provided a comprehensive view of the physical processes that operate in the intracluster medium (ICM), including feedback from the active galactic nucleus (AGN) 3C\,84 and measurements of ICM turbulence. Additionally, the Perseus cluster contains a central radio mini-halo. This diffuse radio source traces cosmic-ray electrons (re-)accelerated in situ in the ICM. 

Here, we report on LOFAR high-band antenna 120--168\,MHz observations of the Perseus cluster that probe a range of four orders of magnitude in angular scales. In our 0.3\arcsec{} (0.11\,kpc) resolution image, we find that the northern extension of the 3C\,84 lobe consists of several narrow 1.5--3\,kpc parallel strands of emission. In addition, we detect steep-spectrum filaments associated with a previous outburst of the central AGN radio emission filling two known X-ray ``ghost'' cavities. 
At 7\arcsec{} resolution (2.6\,kpc), our images show a complex structured radio mini-halo, with several edges and filaments. At resolutions of 26\arcsec{} (10\,kpc) and 80\arcsec{} (29\,kpc), we discover diffuse radio emission with a 1.1\,Mpc extent. We classify this emission as a giant radio halo, and its properties are distinct from the inner mini-halo. We also detect two diffuse sources at projected cluster centric radii of 0.7 and 1.0\,Mpc. 
Finally, we observe a 
0.9\,Mpc trail of radio emission from the cluster member galaxy IC\,310 that connects it to the giant radio halo. Together with other recent studies of relaxed clusters, our LOFAR observations indicate that cluster-wide radio emission could be (more) common in cool-core clusters. In the case of the Perseus cluster, a past off-axis merger event that preserved the cool core might have generated enough turbulence to produce an extended radio halo observable at low frequencies.}


   \maketitle
%

\section{Introduction}

Feedback by active galactic nuclei (AGNs) is an important process that operates in the ICM. It prevents a runaway cooling flow from developing, given the short radiative cooling time of the X-ray-emitting gas in the cluster center \citep[e.g.,][]{2012ARA&A..50..455F,2012NJPh...14e5023M}. The radio AGNs create lobes of relativistic plasma that excavate cavities in the X-ray-emitting ICM, which then rise buoyantly \citep{2000A&A...356..788C}. AGN feedback is ubiquitous in relaxed clusters with short central cooling times \citep[e.g.,][]{2004ApJ...607..800B,2006ApJ...652..216R,2006MNRAS.373..959D}. 

The Perseus cluster \citep[$z=0.0179$;][]{1999ApJS..125...35S} is a well-studied ``cool-core'' galaxy cluster. In the specific case of Perseus, energy from 3C\,84, the radio-loud supermassive black hole associated with the brightest cluster galaxy (BCG) NGC\,1275, is thought to balance the cooling in the cluster core.
X-ray observations have indicated at least two cavity pairs in the cluster core \citep[e.g.,][]{1981ApJ...248...55B,1981ApJ...248...47F,1993MNRAS.264L..25B,2000A&A...356..788C,2003MNRAS.344L..43F,2006MNRAS.366..417F,2011MNRAS.418.2154F}, with the inner pair being co-spatial with the main radio lobes of 3C\,84. High-resolution \textit{Chandra} X-ray observations also show low-amplitude quasi-periodic ripples in the cluster core. These are interpreted as sound waves generated by the AGN and which can balance the radiative cooling within the inner 50\,kpc of the cluster core \citep{2003MNRAS.344L..43F,2017MNRAS.464L...1F}.

Low-frequency observations at 74\,MHz with the Very Large Array (VLA) showed the presence of steep-spectrum radio emission in the direction of the outer older ``ghost cavities'' \citep{2002IAUS..199..189B}. Using  230--470
\,MHz VLA observations, the emission leading toward the northern ghost cavity was resolved into two filamentary structures \citep{2020MNRAS.499.5791G}. However, given the limited sensitivity and resolution of the VLA observations these structures could not be imaged in greater detail. Recent studies suggest that synchrotron-emitting filaments are ubiquitous in the ICM \citep[e.g.,][]{2020A&A...636L...1R,2022ApJ...935..168R,2023MNRAS.523.1933V,2023MNRAS.520.4427M}. It has been proposed that the filaments are bundles of magnetic fields that arise in turbulent magnetohydrodynamics (MHD) flows \citep{2015ApJ...810...93P}. So far, such filaments have not been resolved into individual fibers.
Another explanation for the synchrotron-emitting filaments is given by \cite{2024MNRAS.529L.135G}. This involves time-varying jets that induce fluctuating electric currents within the jets. Then, discharges dissipating the transient patches of electric fields through the filaments illuminate the old jets left by repeated episodes of nuclear activity.

The Perseus cluster is also known to host a 300--400\,kpc radio mini-halo \citep[e.g.,][]{1975A&A....45..223M,1982Natur.299..597N,1990MNRAS.246..477P,1992ApJ...388L..49B,2011A&A...526A...9B,2017MNRAS.469.3872G}. This extended steep-spectrum synchrotron source \citep[$\alpha \approx -1.2$;\footnote{ For the sign convention of the spectral index ($\alpha$), we use $S_\nu \propto\nu^{\alpha}$, with $S_\nu$ being the flux density.}][]{1993PhDT.......392S,2021ApJ...911...56G} surrounds the BCG NGC\,1275. Radio mini-halos are relatively common in massive, relaxed cool-core clusters \citep{2017ApJ...841...71G}. Considering cosmic-ray electron energy losses and the time required to diffuse across a region of more than $\sim100$\,~kpc in size,  radio mini-halos cannot be directly powered by the central radio galaxy. A suggested scenario is that turbulence in the cluster core, for example, induced by gas sloshing motions, can re-accelerate seed cosmic rays \citep[e.g.,][]{2002A&A...386..456G,2008ApJ...675L...9M,2013ApJ...762...78Z}. Alternatively, the synchrotron-emitting cosmic-ray electrons are secondary products from hadronic collisions in the ICM  \citep[e.g.,][]{1980ApJ...239L..93D,2004A&A...413...17P}. The first scenario, re-acceleration by turbulence, has also been invoked to explain radio emission on larger 1--2\,Mpc scales in merging clusters \citep[][]{ 2001ApJ...557..560P,2001MNRAS.320..365B,2007MNRAS.378..245B,2015ApJ...800...60M}, so-called giant radio halos, or even more extended envelopes of emission \citep{2022Natur.609..911C,2022SciA....8.7623B,2024ApJ...961...15N}. In some (partly) relaxed clusters, extended emission on scales larger than those sampled by mini-halos has also been detected \citep{2014MNRAS.444L..44B,2017A&A...603A.125V,2018MNRAS.478.2234S,2021MNRAS.508.3995B,2022A&A...657A..56K}. Most recently, \cite{2024A&A...686A..82B} found that that cluster-scale diffuse radio emission is not present in all cool-core clusters when observed with LOFAR, and it is correlated with the presence of cold fronts. The idea is that the presence of cluster-scale diffuse radio emission indicates that these clusters are not fully relaxed. Until now, no evidence has been found in the Perseus cluster for cluster-scale diffuse emission, and whether cluster-wide emission beyond the scales of mini-halos is ubiquitous in cool-core clusters remains
an open question. This is because current studies are not yet deep enough to probe well below the radio power--mass relation in these systems \citep{2024A&A...686A..82B}.

The Perseus cluster is the brightest X-ray cluster in the sky and the closest example of a cool-core system. Therefore, it serves as an important test bed to study radio-mode feedback, steep-spectrum synchrotron-emitting filaments, and extended diffuse radio emission in non-merging clusters. By observing at low frequencies, we are sensitive to steep-spectrum emission from past AGN outbursts and diffuse radio emission. Moreover, studies indicate that the diffuse emission beyond the cores of (semi-)relaxed clusters has a steeper spectral index than radio mini-halos \citep[e.g.,][]{2017A&A...603A.125V,2018MNRAS.478.2234S,2021MNRAS.508.3995B,2023A&A...678A.133B,2024A&A...686A..44R}. However, similarly to some other nearby cool-core clusters, the bright central radio AGN 3C\,84 poses major challenges, overpowering possible faint extended emission.

It is important to determine the spectral index of the giant radio halo in Perseus. 
Studies indicate that the diffuse emission beyond the cores of (semi-)relaxed clusters has a steeper spectral index \citep[e.g.,][]{2017A&A...603A.125V,2018MNRAS.478.2234S,2021MNRAS.508.3995B,2023A&A...678A.133B,2024A&A...686A..44R} than radio mini-halos.

In this work, we present  International LOFAR Telescope \citep{vanhaarlem+13} high-band antenna (HBA) observations of the Perseus cluster, including baselines of up to 2,000\,km. The outline of this paper is as follows. In Sect.~\ref{sec:datareduction}, we describe the observations and data reduction. The results are presented in Sect.~\ref{sec:results}. This is followed by a discussion and conclusions in Sects.~\ref{sec:discussion} and ~\ref{sec:conclusion}. We assume a $\Lambda$CDM cosmology with $H_{0} = 70$~km~s$^{-1}$~Mpc$^{-1}$, $\Omega_{m,0} = 0.3$, and $\Omega_{\Lambda,0} = 0.7$. At the distance of Perseus, 1\arcsec{} corresponds to 0.366\,kpc.

\begin{figure*}[h!]
\centering
\includegraphics[width=0.9\textwidth]{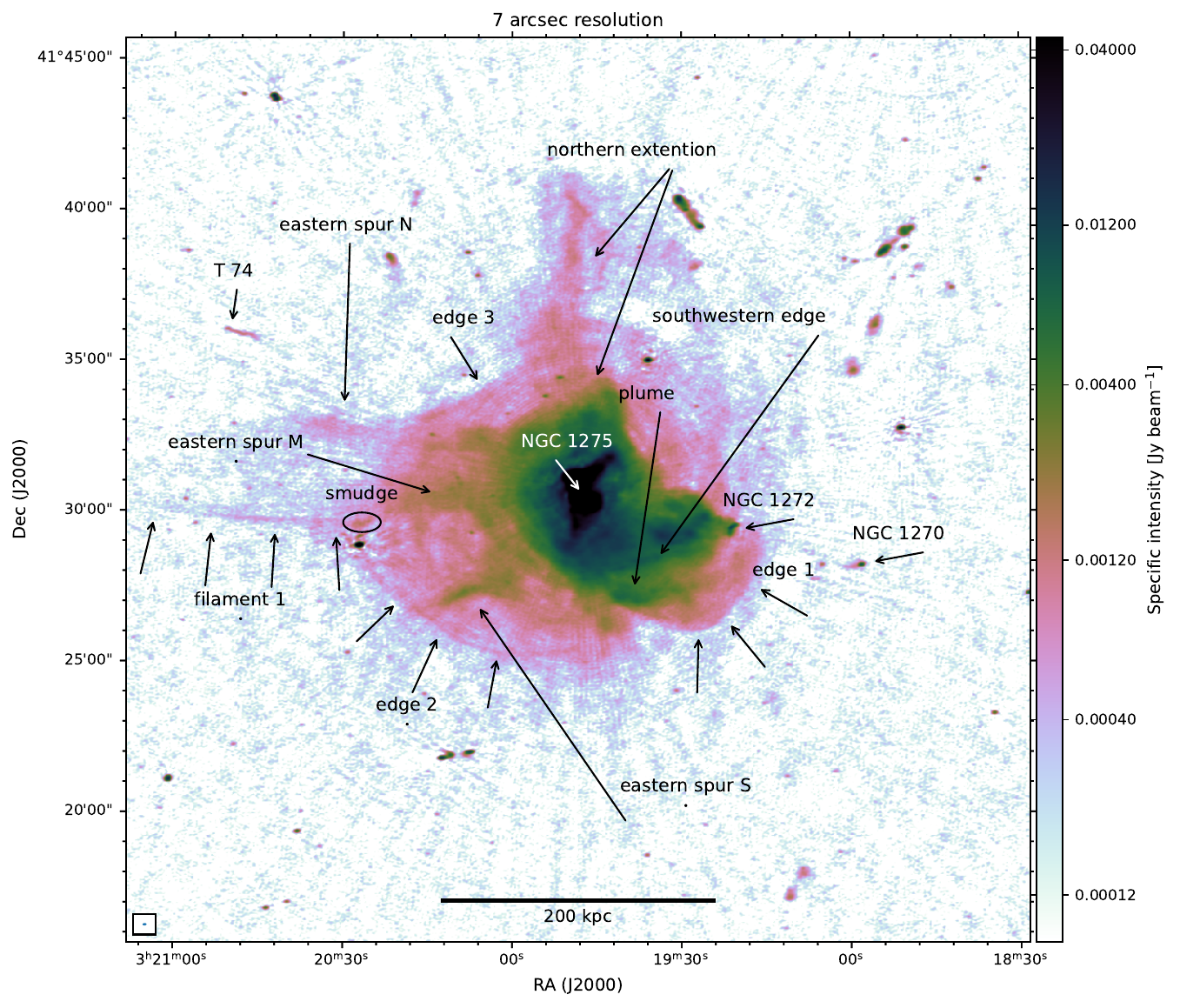}
\caption{144\,MHz image of the central region of the Perseus cluster at 7\arcsec{} resolution. The logarithmic color scale ranges between 1 and 500 times the r.m.s. noise of 98\,$\mu$Jy\,beam$^{-1}$. Various features are labeled, starting from those in \cite{2017MNRAS.469.3872G}. The beam size is indicated in the bottom-left corner.}
\label{fig:HBAfullzoom}
\end{figure*}

\section{Observations and data reduction}
\label{sec:datareduction}
For this work, we used the pointing P049+41 (SAS Id: 557730) taken as part of the LOFAR Two-metre Sky Survey \citep[LoTSS;][]{shimwell+19,2022A&A...659A...1S} in the 120--168\,MHz frequency range with the LOFAR HBA. The observation was obtained on Nov 3, 2016, and centered on RA=49.247\degr{} and Dec=41.380\degr{}. Additionally, we used archival LOFAR data (L922461, SAS Id: 691332, PI: Bempong-Manful) from Dec 14, 2018,  with an identical frequency setup to the LoTSS observation. L922461 is centered on RA=49.565\degr{} and Dec=41.858\degr{}. L922461 was only used for {high-resolution} imaging {(0.3\arcsec--3.4\arcsec)} of 3C\,84. These data had 11 international LOFAR stations participating in the observations compared to the 9 from P049+41. Both 8\,hr observations were bracketed by two 10~min scans on primary calibrators. The data reduction follows the steps described in \cite{2022A&A...668A..65T,2022A&A...658A..44R}. First, radio frequency interference was removed with the \texttt{AOFlagger} \citep{2010MNRAS.405..155O,2012MNRAS.422..563O,2013A&A...549A..11O} and instrumental effects were corrected for using the \texttt{prefactor} \citep{vanweeren+16,williams+16,degasperin+19} pipeline. We then ran the \texttt{lofar-vlbi} pipeline \citep{2022A&A...658A...1M} to create a dataset centered on 3C\,84. This dataset was self-calibrated using the facet calibration scheme described in \cite{2021A&A...651A.115V}, employing a simple four-component starting model for 3C\,84 (a single
point source plus three Gaussians). 

To image the full cluster field with the Dutch baselines, we needed to remove the bright compact core of 3C\,84 as it otherwise severely limits the dynamic range. We did this by constructing a high-resolution 0.3\arcsec{} model of the source, removing baselines shorter than 40\,k$\lambda$. We subtracted this model from the visibility data with the obtained calibration solutions. This dataset was then processed employing the \texttt{ddf-pipeline} \cite{2021A&A...648A...1T}, which utilizes \texttt{killMS} \citep{tasse14,smirnov+tasse15} and \texttt{DDFacet} \citep{tasse+18}. Finally, to improve the dynamic range in the cluster region further -- because additional bright emission is present beyond the core of 3C\,84 -- we imaged the central ${\sim}2\degr$ using \texttt{WSClean} in facet-mode (10 facets) employing multi-scale deconvolution \citep{offringa+14,2017MNRAS.471..301O}. Here, the facet solutions were determined with the method described in \cite{2022A&A...668A.107D}. To study the radio mini-halo, tailed radio galaxies, and other extended sources we made images with various weighting schemes, resulting in beam sizes of 8\arcsec$\times$5\arcsec, 26\arcsec$\times$25\arcsec, and 90\arcsec$\times$70\arcsec{}; we employed a Gaussian uv taper for the latter two images. No inner uv cut was applied during the imaging to allow a search for very extended cluster emission,  resulting in a shortest (projected) baseline of 40\,m, corresponding to an angular scale of about 3.5\degr{}. Using the noise level of $98$\,$\mu$Jy\,beam$^{-1}$ in the 8\arcsec$\times$5\arcsec{} resolution image and the original peak flux of 3C\,84's core (at the same resolution), we compute a dynamic range of $8\times10^{4}$.

To study 3C\,84 on small spatial scales, we phased up the LOFAR core stations into a larger virtual station to speed up the imaging process and suppress emission from other sources within the primary beam. The {high-resolution} images {(0.3\arcsec--3.4\arcsec)} presented in this paper are from the L922461 pointing as this dataset has better uv coverage due to the higher number of international stations participating in the observation. The high-resolution image of P049+41 has already been presented in \cite{2022A&A...668A..65T}. Our new L922461 sub-arcsecond resolution image  ($0.41\arcsec \times 0.23\arcsec$) has a noise level of 76\,$\mu$Jy\,beam$^{-1}$, compared to the  130\,$\mu$Jy\,beam$^{-1}$ from {\cite{2022A&A...668A..65T}}. In Table~\ref{tab:imageproperties}, we give an overview of images produced and their properties such as beam size and r.m.s. noise level.

\begin{figure}[h!]
\centering
\includegraphics[width=0.49\textwidth]{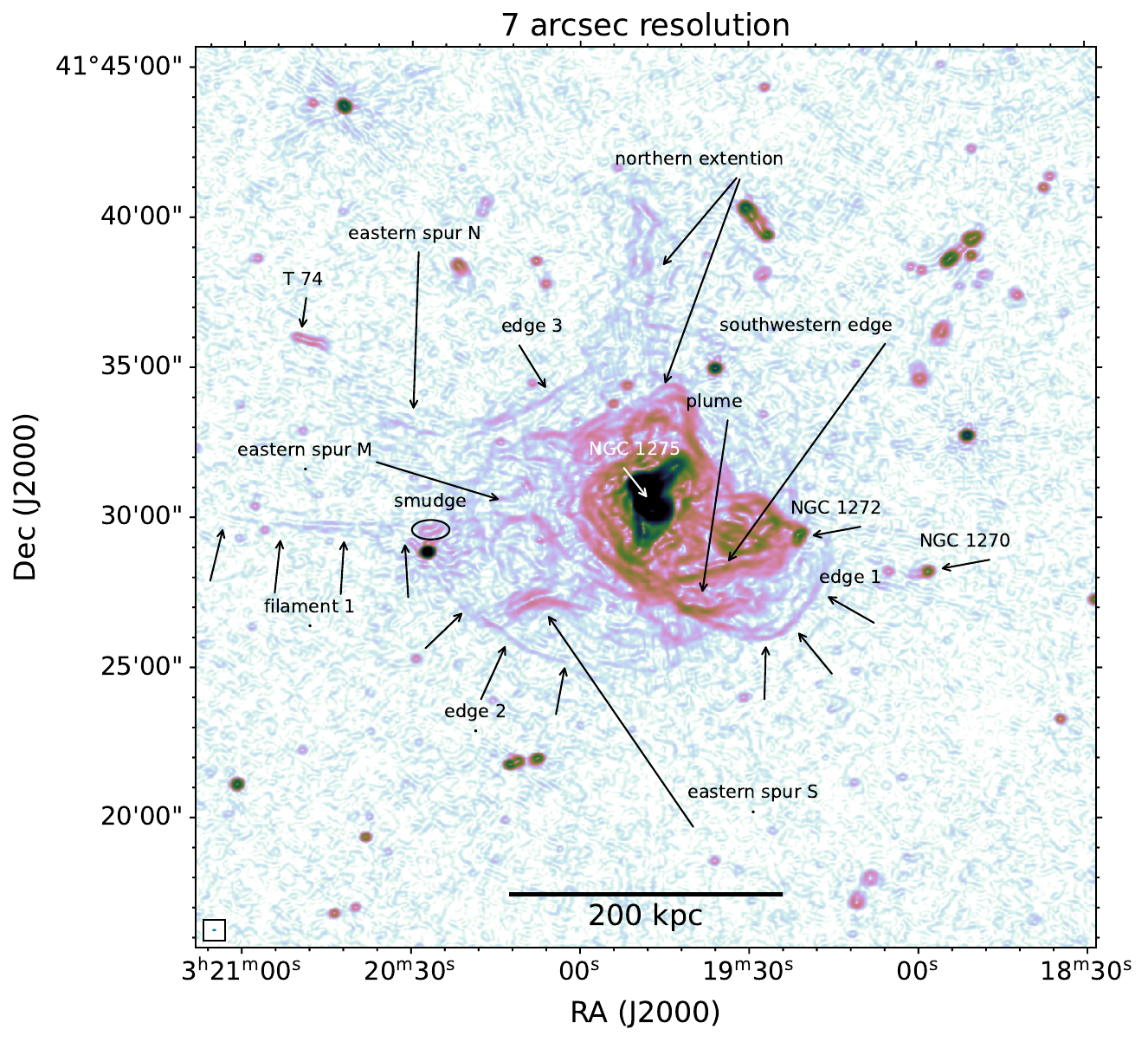}
\caption{{144\,MHz image of the center of the Perseus cluster at 7\arcsec{} resolution, processed with a GGM filter with $\sigma=3.75\arcsec{}$, highlighting small-scale features in the radio mini-halo region.}}
\label{fig:HBAfullzoomGGM}
\end{figure}

\begin{figure*}[h!]
\centering
\includegraphics[width=0.95\textwidth]{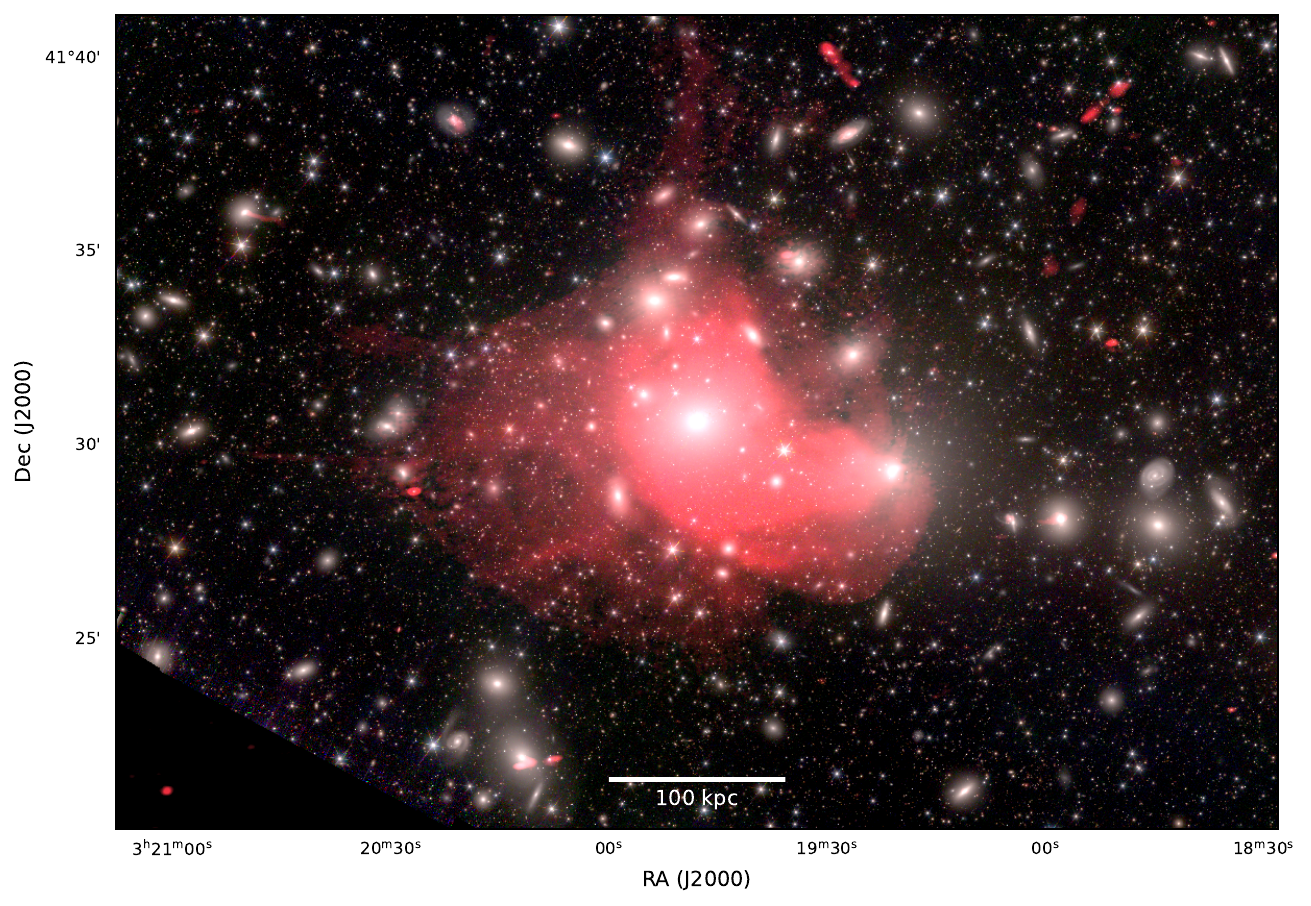}
\caption{Euclid Near-Infrared Spectrometer and Photometer \citep[NISP;][]{2024arXiv240513493E,2024arXiv240513491E} image of the Perseus cluster from \cite{ERO}; for more details on these data, see \cite{2024arXiv240513496C,2024arXiv240513501C}. The $Y_E$-, $J_E$-, and $H_E$ -band images were used for the blue, green, and red channels. The 144\,MHz LOFAR image at 7\arcsec{} resolution is overlaid in red.}
\label{fig:HBAEuclid}
\end{figure*}

\begin{table*}[h!]
\
\begin{center}
\caption{LOFAR image properties.}
\begin{tabular}{llllll} 
\hline
\hline
Image name &Stations & Robust weighting & uvtaper               & Resolution, position angle  & Sensitivity\\
&        && \arcsec{}    & \arcsec{} $\times$ \arcsec{}, \degr{} & $\mu$Jy\,beam$^{-1}$ \\
\hline
7\arcsec &Dutch & $-0.5$        & \ldots  &$8.2\times4.8$ , 93 & 98 \\
26\arcsec &Dutch &  $-0.5$  & 20        & $26\times25$  , 8 & 337 \\
80\arcsec &Dutch & $-0.5$ & 60  & $90\times70$ , 70 & 951 \\
0.3\arcsec&All &  $-1.25$  & \ldots & $0.41\times0.23$,  3 & 76\ \\
1.6\arcsec &All & $-1.25$& 1.0  & $2.4\times0.8$,  108 & 140\ \\
3.4\arcsec &All & $-1.25$ &1.5  & $4.0\times2.7$,  103 & 370\ \\
\hline
\end{tabular}\label{tab:imageproperties}
\end{center}
\end{table*}

\subsection{Spectral index maps and compact source subtraction} 
To make spectral index maps, we employed the 
VLA 230--470\,MHz P-band A-array and the 1--2\,GHz L-band C+D-array images from \cite{2021ApJ...911...56G}. 
We re-imaged the Dutch baseline LOFAR data using inner-uv cuts corresponding to the shortest baselines in the VLA data. Images were convolved to the same resolution and re-gridded on a common pixel grid utilizing \texttt{CASA}  \citep{2007ASPC..376..127M,2022PASP..134k4501C} before computing the spectral index. For the spectral index uncertainty, we took both the r.m.s. map noise and uncertainty of the absolute flux scale into account. We adopted an absolute flux-scale uncertainty of 10\%, 5\%, and 2.5\% for the HBA, VLA P-band, and L-band, respectively \citep{2016ApJ...818..204V,2017ApJS..230....7P,
2022A&A...659A...1S}.

We also produced source-subtracted images, taking out the contribution from the more compact radio sources. This was done by first identifying the location of compact sources using \texttt{PyBDSF} \citep{2015ascl.soft02007M} on the 7\arcsec{} resolution image. We manually inspected the source list and removed some components that are associated with more extended emission (e.g., from sharp ``edges'' in the mini-halo).  The clean components, corresponding to the location of the detected compact sources, were then subtracted from the uv data. Because of the complexity of the radio mini-halo, 3C\,84's lobes, various tailed AGNs (e.g., IC\,310, NGC\,1265, and NGC\,1272), and the range of spatial scales involved, we decided not to remove these sources. The data were then re-imaged with the same settings we used to make the 26\arcsec{} and 80\arcsec{} low-resolution images (see Table~\ref{tab:imageproperties}).

\begin{figure*}[t!]
\centering
\includegraphics[width=0.49\textwidth]{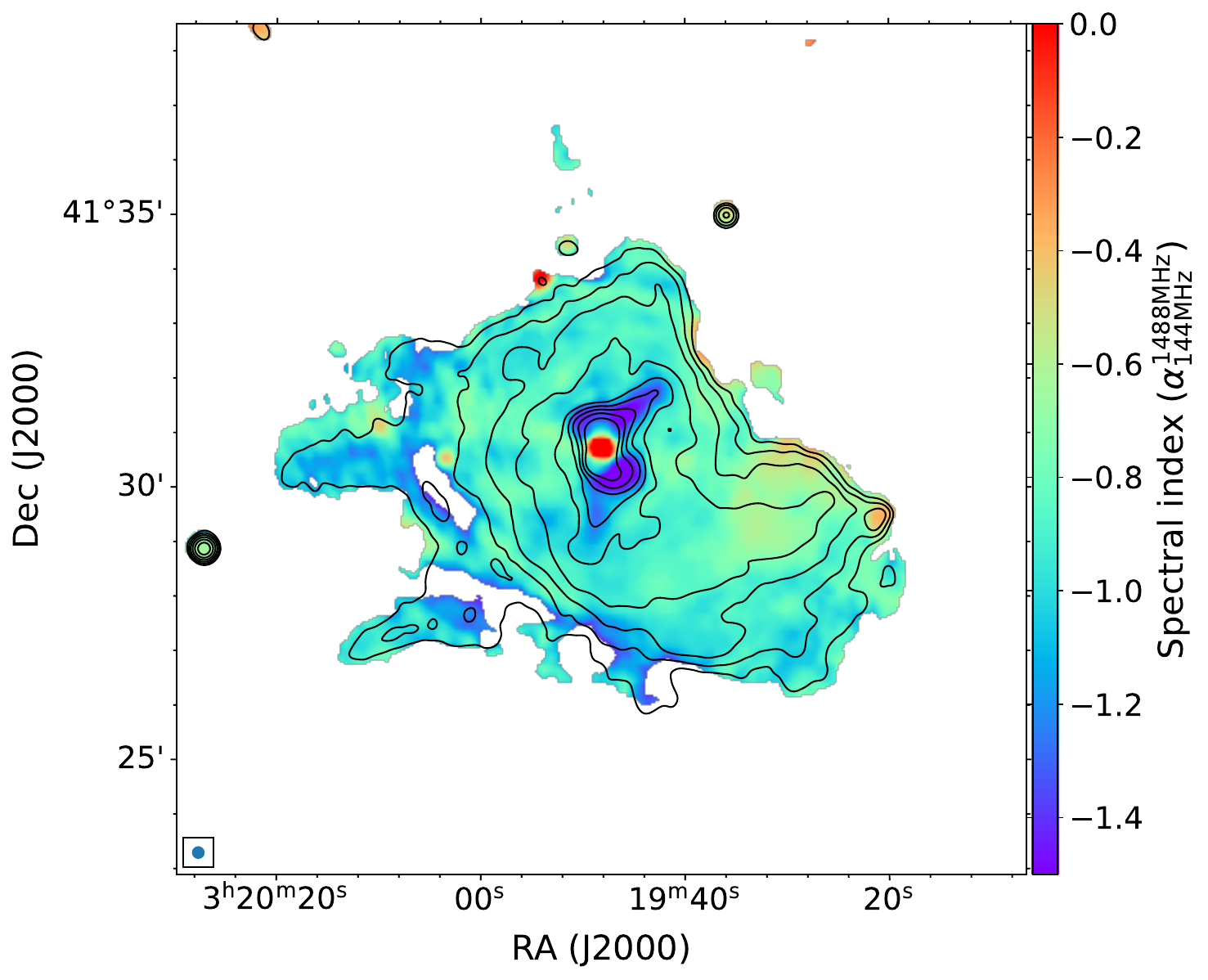}
\includegraphics[width=0.49\textwidth]{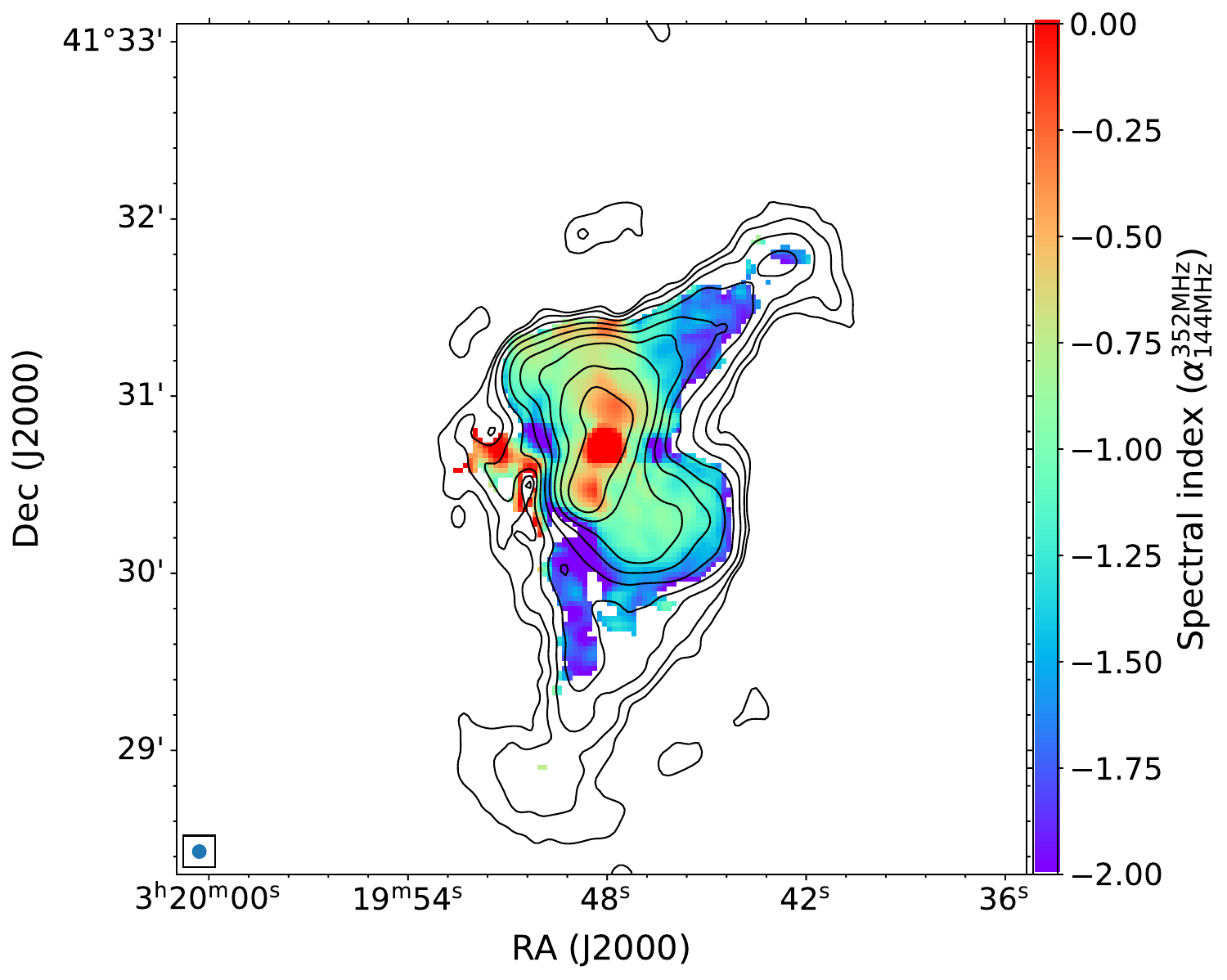}
\caption{Radio spectral index maps between 144 and 1488\,MHz at 14\arcsec{} resolution (left panel) and  between 144 and 352\,MHz at 5\arcsec{} resolution (right panel). The beam sizes are indicated in the bottom left corners. Pixels below $3\sigma_{\rm{rms}}$ and $8\sigma_{\rm{rms}}$ were blanked in the $\alpha_{144}^{1488}$ and $\alpha_{144}^{352}$-maps, respectively. In both maps, the radio contours come from the 144\,MHz images and are drawn at levels of $5\sigma_{\rm{rms}} \times [1,2,4,8,\ldots]$. The corresponding spectral index uncertainty maps are shown in Fig.~\ref{fig:spixerr}.} 
\label{fig:spix}
\end{figure*}

\begin{figure*}[t!]
\centering
\includegraphics[width=0.49\textwidth]{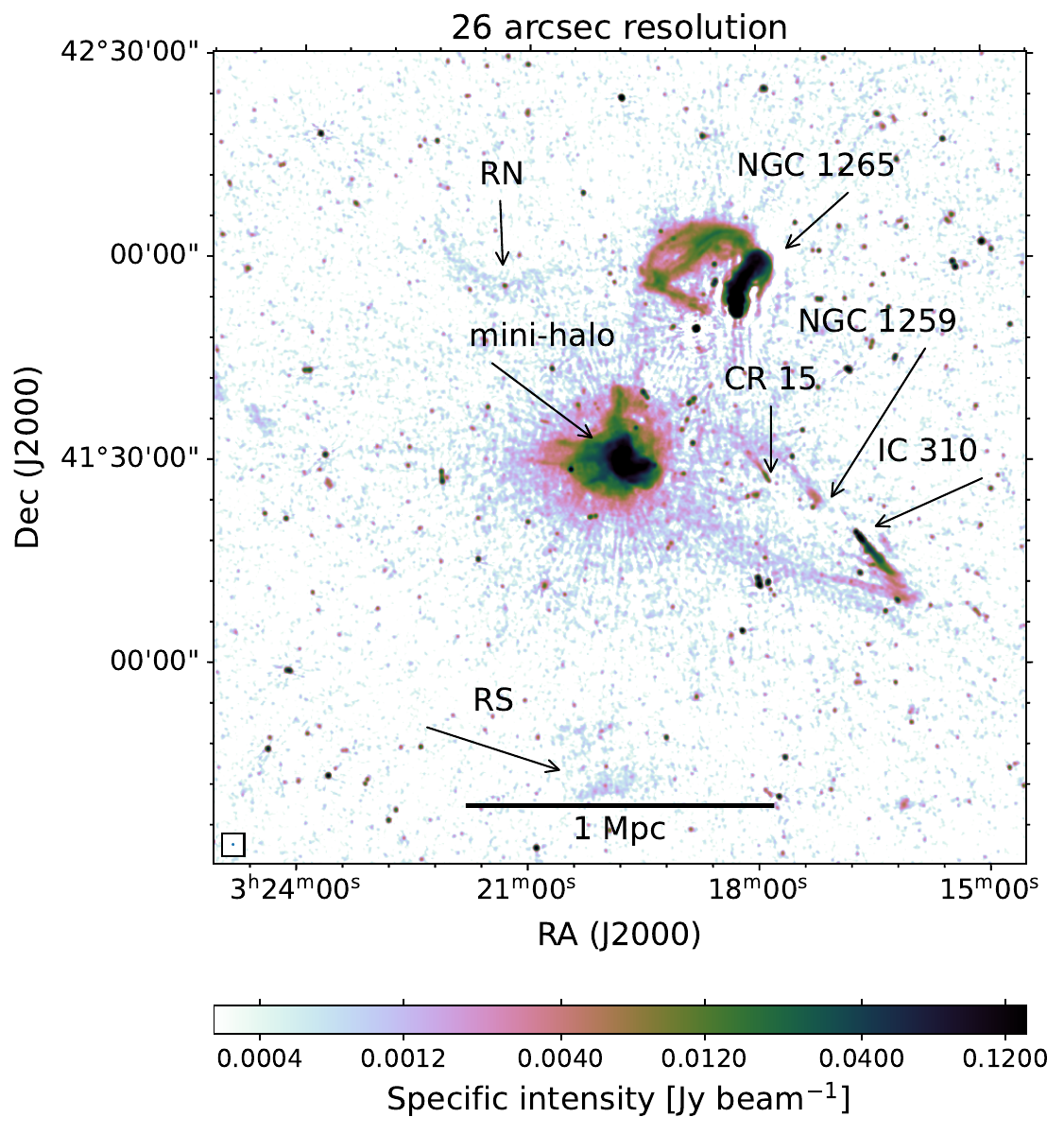}
\includegraphics[width=0.49\textwidth]{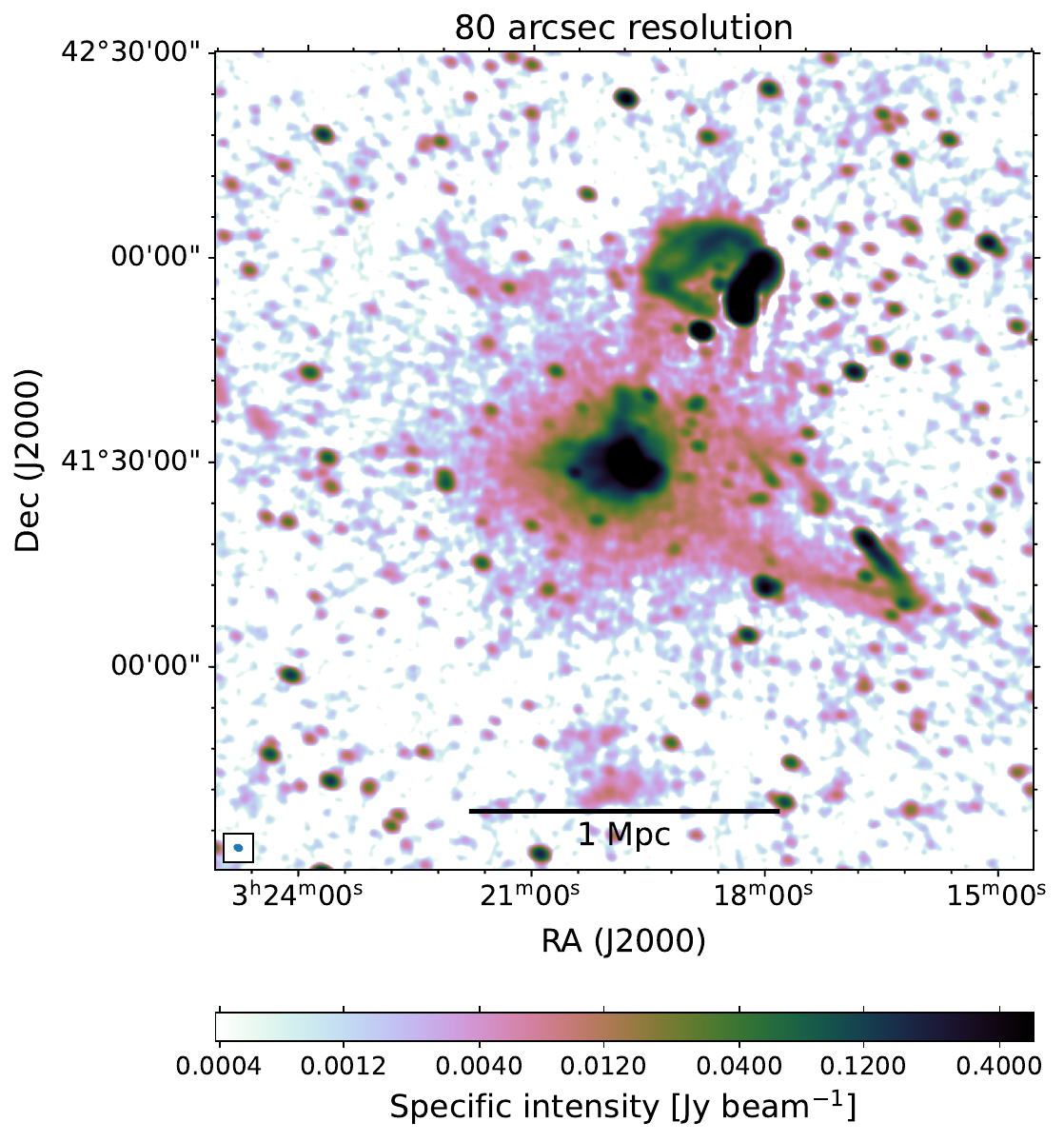}
\caption{144\,MHz images of the Perseus cluster at a resolution of 26\arcsec{}(left) and 80\arcsec{}(right). The beam sizes are indicated in the bottom left corners. Several extended radio sources are labeled.}
\label{fig:HBAlowres}
\end{figure*}

\subsection{Chandra observations}

The Perseus cluster has been the target of multiple \textit{Chandra} observations. Here, we make use of the same data presented in \cite{2022A&A...668A..65T}, which resulted in a mosaic image of the cluster center with a net exposure of 1.4~Ms (ObsIDs: 1513, 3209, 4289, 4946\dots4953, 6139, 6145, 6146, 11713\dots11716, 12025, 12033, 12036, 12037). This was obtained by merging the clean event files of the different ObsIDs to produce an exposure-corrected image in the 0.5--2.0~keV energy band. Following \cite{2011MNRAS.418.2154F}, we produced a fractional residual image of the cluster's X-ray emission to highlight substructures in the ICM.

\section{Results}
\label{sec:results}
Below, we describe the results obtained from the LOFAR HBA observations. We separate this section into one subsection on the extended radio emission (Sect.~\ref{sec:extended}) and one on the radio galaxies, including 3C\,84 (Sect.~\ref{sec:tails}). A summary of the properties of the observed sources is given in Table~\ref{tab:sourceproperties}.

\begin{table}[h!]
\
\begin{center}
\caption{Source properties.}
\begin{tabular}{lll} 
\hline
\hline
Source & Flux density & projected length$^{a}$  \\
   & Jy & kpc \\
\hline
mini-halo & $20.5 \pm 0.2$ & 380\\
giant halo & $5.8\pm 0.6$ & 1100 \\
relic north (RN) &$0.19\pm 0.02$ & 450\\  
relic south (RS) & $0.22 \pm 0.02$ & 450\\
NGC\,1272$^{b}$ & \ldots & \ldots \\
3C\,84$^{c}$ &  \ldots & 80 \\
T\,74 & $\left(14.8\pm0.2\right) \times 10^{-3}$ & 24\\
NGC1259 (CR\,10) & $0.16\pm0.02$ & 250 \\
CR\,15 & $0.12\pm0.01$ & 130 \\
NGC\,1270 & $\left(19.9\pm0.2\right) \times 10^{-3}$ & 14 \\
IC\,310  & $2.1\pm0.2$ & 700 \\
NGC\,1265 & $28\pm3$ & 420  \\
\hline
\hline
\end{tabular}\label{tab:sourceproperties}
\end{center}
$^{a}$ Synonymous with "linear size"; see \cite{2023A&A...672A.163O}, footnote 1. \\
$^{b}$ Blended with radio mini-halo.\\
$^{c}$ The radio core was removed from the uv data; the projected length measurement was taken based on the radio emission associated with the northern and southern ghost cavities.
\end{table}

\subsection{Extended radio emission}
\label{sec:extended}

The LOFAR 7\arcsec{} resolution image of the cluster central region is displayed in Fig.\ref{fig:HBAfullzoom}. We label various features on this image. A combined LOFAR--Euclid image of the central region of the cluster is shown in Fig.~\ref{fig:HBAEuclid}.
The main structure visible in the 144\,MHz image is the radio mini-halo. It displays a complex morphology, with several surface brightness edges and narrow filaments of emission. {To better highlight some of these features, we created a Gaussian gradient magnitude (GGM) filtered image using {\tt scipy} \citep{2020SciPy-NMeth}, which is shown Fig.~\ref{fig:HBAfullzoomGGM}. GGM images measure the gradient in the surface-brightness distribution and help to bring out sharp features on a chosen scale \citep[e.g.,][]{2016MNRAS.460.1898S,2016MNRAS.461..684W}.} 

On large scales, the mini-halo emission is somewhat asymmetric, with the mini-halo being more extended to the east. There are also spurs of emission extending to the north and east, labeled as the ``northern extension'' and three ``eastern spurs'' (S, M, and N). ``Eastern spur M'' appears to connect to a remarkable long narrow filament (``filament 1'') pointing to the east. This filament is 170\,kpc long and has a width of about 4 \,kpc. The filament has no clear optical counterpart but we note an east-west elongated ``smudge'' of emission near the western end of the filament. This smudge seems associated with the cluster member galaxy \object{PER\,163} \citep{1978AJ.....83..732S}, with the galaxy being located on the eastern side of the smudge. However, the connection to filament 1 remains unclear.   
We also detect three edges (``edge 1'', ``edge 2'', and ``edge 3'') located near the outer visible boundary of the radio mini-halo. 

The spectral index map between 144\,MHz and 1.5\,GHz is displayed in Fig.~\ref{fig:spix} (left panel). The mini-halo has a relatively constant spectral index ranging between $-0.9$ and $-1.1$, which is generally consistent with previous work \citep{1993PhDT.......392S,2021ApJ...911...56G} given that the frequency ranges are not identical and there are uncertainties in the absolute flux scales. The map shows a slightly flatter ($-0.6$ to $-0.8$) spectral index in the direction of NGC\,1272 (which would be expected if some of the lobe emission from NGC\,1272 is blended with that of the mini-halo). This flatter region was also seen by \cite{2021ApJ...911...56G}.
The uncertainties for these spectral index values are about 0.05 to 0.07 (see Fig.~\ref{fig:spixerr}, top panel), so this flattening is significant. There is a hint of steeper spectral index values ($-1.2\pm0.1$) near the southeastern boundary of the radio mini-halo and for the eastern spur.

\begin{figure}[t!]
\centering
\includegraphics[width=0.49\textwidth]{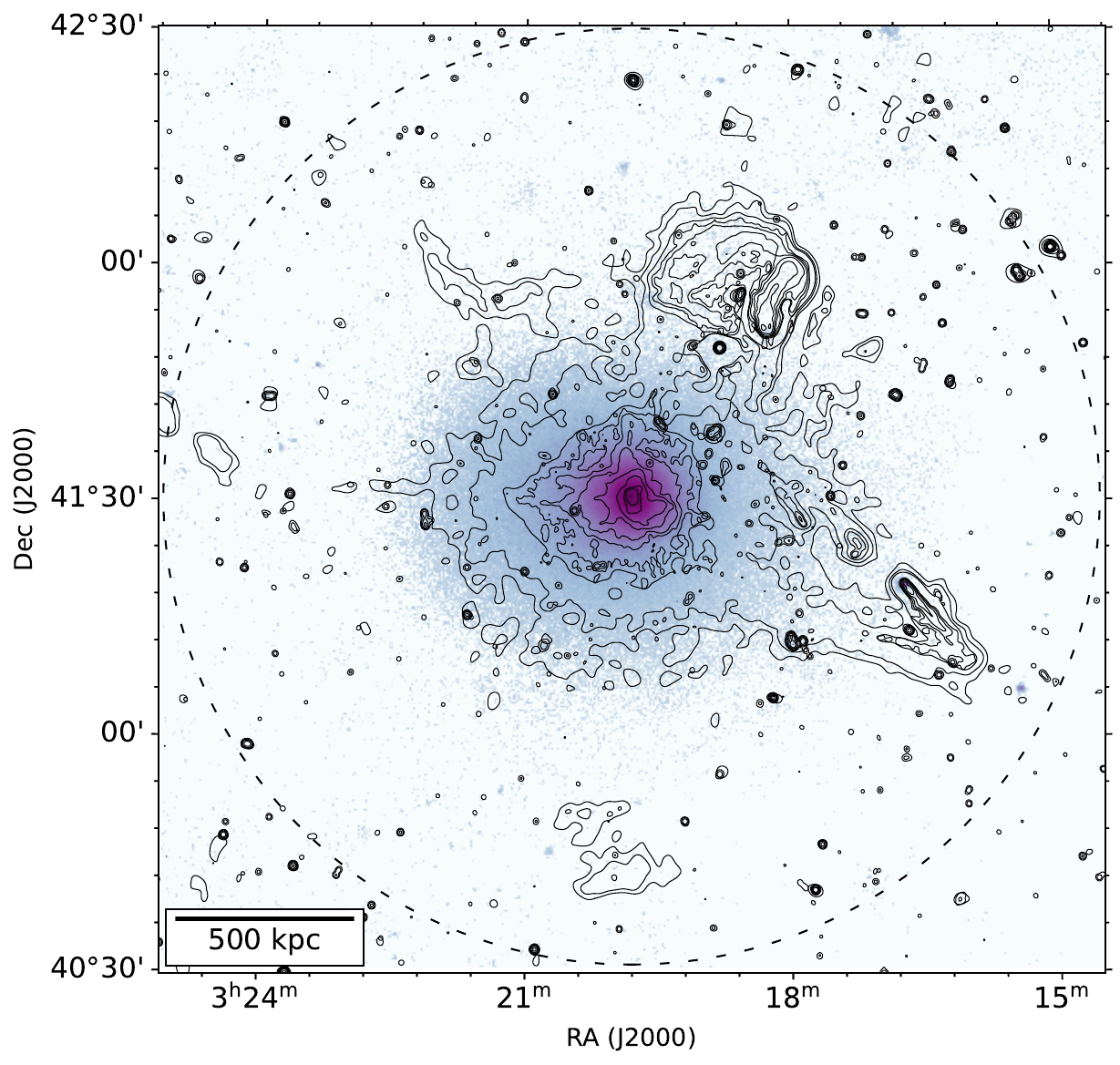}
\caption{X-ray ROSAT PSPC 0.1--2.4\,keV image \citep{1999A&A...349..389V} overlaid with 144 MHz LOFAR radio contours. The radio contours are from the 26\arcsec{} and 80\arcsec{}   resolution images, with the emission from compact sources subtracted in the latter one. Radio contours are logarithmically spaced, starting at $5\sigma_{\rm{rms}}$ and $2.5\sigma_{\rm{rms}}$, respectively. The dashed circle indicates $R_{500}$ \citep{2014MNRAS.437.3939U}.}
\label{fig:PSPC}
\end{figure}

In the 26\arcsec{} resolution LOFAR image shown in Fig.~\ref{fig:HBAlowres} (left panel), the mini-halo ``spurs'' are also clearly detected. In addition, faint radio emission is visible beyond the boundary of the mini-halo as seen in the 7\arcsec{} resolution image (Fig.~\ref{fig:HBAfullzoom}). This faint emission is detected at a higher signal-to-noise ratio in the 80\arcsec{} resolution image (see Fig.~\ref{fig:HBAlowres}, right panel). It fully envelopes the mini-halo and has a projected size of 1.1\,Mpc. The 26\arcsec{} and 80\arcsec{} low-resolution images, with the emission from compact sources subtracted, are shown in Fig.~\ref{fig:HBAlowressub}. In Fig.~\ref{fig:PSPC}, we overlay the low-resolution radio maps on a ROSAT PSPC 0.1--2.4\,keV X-ray image. The radio emission at 80\arcsec{} resolution traces a large part of the X-ray-emitting ICM visible in the ROSAT PSPC image. We classify the extended low-surface-brightness radio emission surrounding the radio mini-halo as a giant radio halo. This agrees with the projected size of 1.1\,Mpc and the fact the radio emission is co-located with the X-ray emission from the ICM.  The low-resolution radio images also reveal a morphological connection between the emission of the tailed radio galaxies IC\,310, CR\,15, and NGC\,1259 and the giant radio halo. The results for the radio galaxies are described in Sect.~\ref{sec:tails}.

\begin{figure}[t!]
\centering
\includegraphics[width=0.5\textwidth]{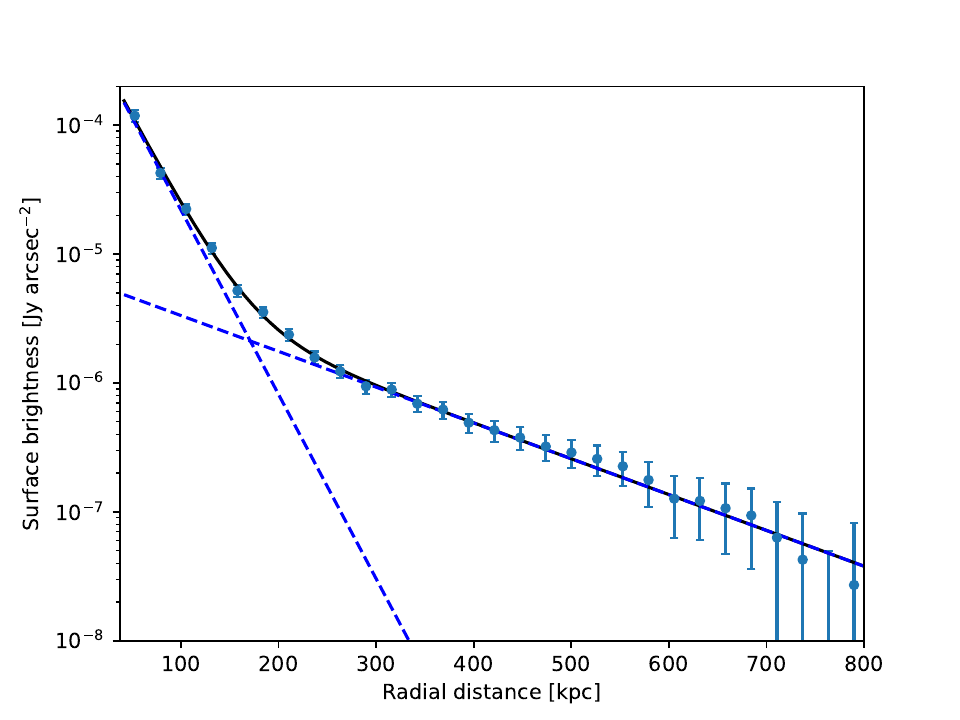}
\vspace{-6mm}
\caption{Radial halo surface-brightness profile at 144\,MHz. Data points show the surface brightness extracted in 26\,kpc annuli centered on NGC\,1275. The profile is well fit by a double exponential model (solid black line), with the blue dashed lines showing the individual exponential components. The inner region below 40\,kpc, which is contaminated by 3C\,84, is not shown.} 
\label{fig:radprofile}
\end{figure}

To further characterize the radio mini-halo and outer giant halo, we extracted a radial surface-brightness profile. For this we used the compact source-subtracted image at 80\arcsec{} resolution (see Fig.~\ref{fig:HBAlowressub}, right panel). In addition, we masked the remaining extended emission not belonging to the radio (mini-)halo. These masked regions are indicated in red in this figure. The resulting profile (see Fig~\ref{fig:radprofile}) shows two very distinct exponential components.  An inner component with a steep slope corresponding to the previously known radio mini-halo and an outer fainter component with a shallower slope.

Given the shape of
the profile in Fig~\ref{fig:radprofile} and the fact that radio halos and mini-halos have commonly been modeled with exponential profiles \citep{2009A&A...499..679M,2021A&C....3500464B}, we fit a double exponential model of the form 
\begin{equation}
I_{\nu}\left(r\right) = I_{\nu,\rm{0,inner}} e^{r/r_{\rm{e,inner}}} +  I_{\nu,\rm{0,outer}} e^{r/r_{\rm{e,outer}}} \mbox{ .}
\label{eq:profile}
\end{equation}
We obtain $I_{\nu,\rm{0,inner}} = 590\pm61$\,$\mu$Jy\,arcsec$^{-2}$, $I_{\nu,\rm{0,outer}} = 6.3\pm0.7$\,$\mu$Jy\,arcsec$^{-2}$, $r_{\rm{e,inner}}=30.4\pm0.2$\,kpc, and $r_{\rm{e,outer}}=156\pm2$\,kpc.
Comparing the data with the model, we see that it gives a good description of the observed profile. It is also worth noting that the transition between the halo components happens at a distance of about $0.2R_{500}$ \citep{2014MNRAS.437.3939U}\footnote{$R_{500}= 59.6\arcmin$ (1.31\,Mpc).}, which was also considered by \cite{2017ApJ...841...71G} as the maximum extent of mini-halos. From this fit, we obtain the integrated flux densities of the mini-halo (inner) and giant radio halo (outer) components integrating from $r=0$ to $3r_e$ \citep[as has been done in][]{2021A&A...651A.115V,2022A&A...660A..78B}, giving 80\% of the flux density compared to integrating to $r=\infty$. This gives $S_{\rm{144, inner}} =20.5 \pm 2.1$\,Jy and $S_{\rm{144, outer}} =5.8 \pm 0.6$\,Jy. These  ``$3r_e$ flux densities'' correspond to 150\,MHz radio powers of $(1.42\pm0.15) \times 10^{25}$ and $(4.0\pm0.5)\times 10^{24}$~W~Hz$^{-1}$, respectively. To compute the radio power, we adopted a spectral index of $-1.1$ (based on Fig.~\ref{fig:spix}) for the inner component and $-1.3$ for the outer one.\footnote{$-1.3$ is a typical value for giant radio halos \citep[e.g.,][]{vanweeren+19}.} Given the low redshift of the cluster and central observing frequency of 144\,MHz, the 150\,MHz power does not strongly depend on the adopted value of the spectral index. The advantage of using Eq.~\ref{eq:profile} to obtain the flux density is that we can ``cleanly'' separate the contribution of the radio mini-halo from the giant halo and extrapolate across the masked regions. Combining the integrated flux density of both components, this method, however, agrees quite well with a manual measurement of 25.2\,Jy across the $3\sigma_{\rm{rms}}$ contour extent of the full halo emission. Here, we simply subtracted the flux densities of the embedded emission from 3C\,84 and NGC\,1272 and the extended IC\,310 trail.\footnote{Again using the source-subtracted image at 80\arcsec{} resolution; see Fig.~\ref{fig:HBAlowressub} (right panel).}

To the north and south, at projected cluster centric radii of 0.7 and 1.0\,Mpc, 
two irregular-shaped diffuse sources are detected. These are labeled RN and RS in Fig.~\ref{fig:HBAlowres} (left panel). 
The sources have projected lengths of about 400 to 500\,kpc, with the northern source having a concave shape and the southern one consisting of two substructures. No optical counterparts are identified for these sources. Given their projected lengths, lack of optical counterparts, and location in the cluster periphery, we classify these two sources as radio relics. We use the compact source-subtracted image at 80\arcsec{} resolution to measure the integrated flux densities of the radio relics in the regions indicated in Fig.~\ref{fig:HBAlowressub}, obtaining $S_{\rm{144,RN}}=0.19\pm 0.02$\,Jy and  $S_{\rm{144,RS}}=0.22\pm 0.02$\,Jy. This corresponds to 150\,MHz radio powers of $(1.3\pm0.1)\times 10^{23}$ and $(1.5\pm0.2) \times 10^{23}$~W~Hz$^{-1}$, respectively. To compute the radio power we adopted a spectral index value of $-1.1$ \citep[e.g.,][]{vanweeren+19}.


\subsection{3C\,84 and tailed radio galaxies}
\label{sec:tails}

\begin{figure*}[t!]
\centering
\includegraphics[width=0.48\textwidth]{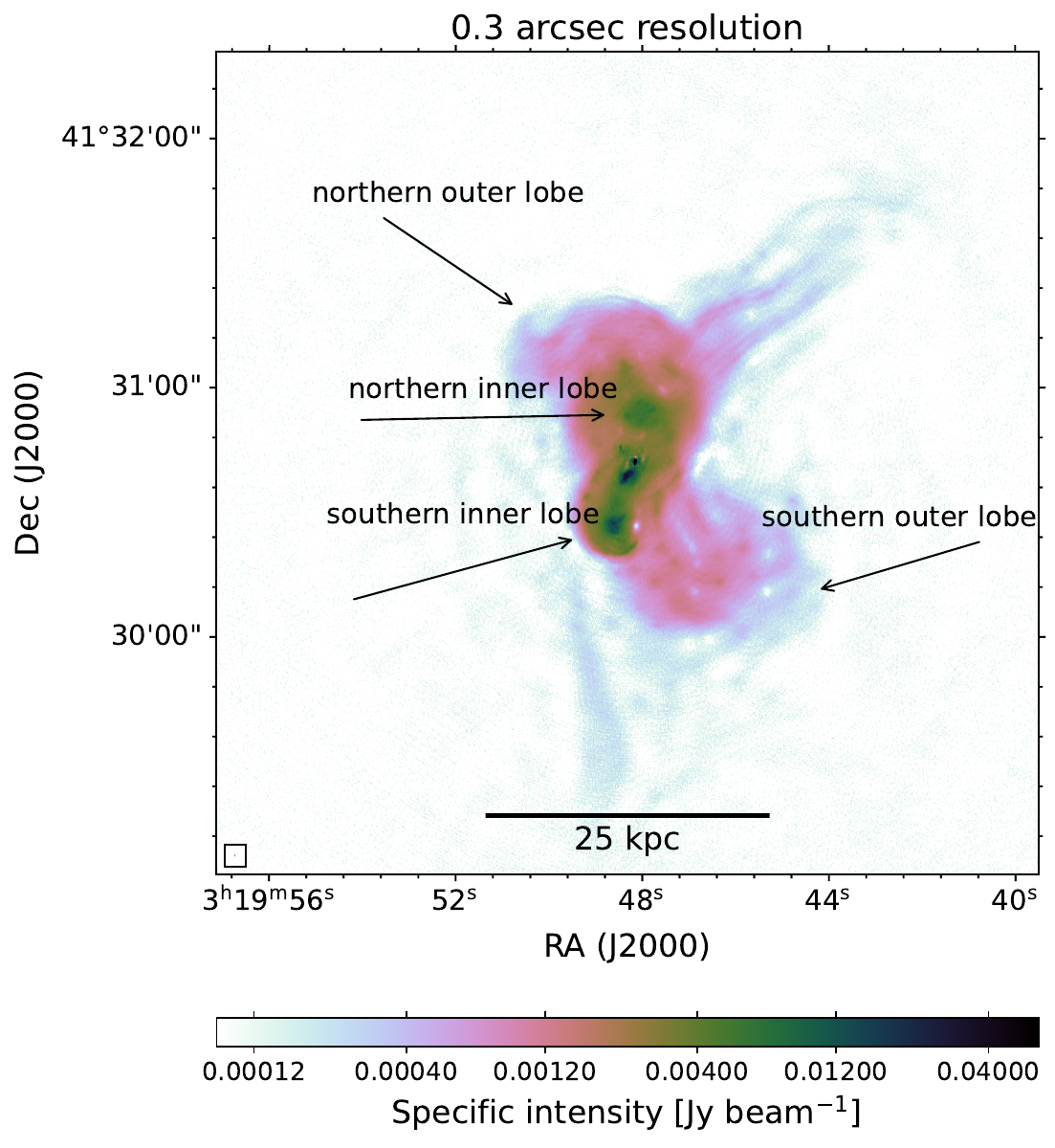}
\includegraphics[width=0.50\textwidth]{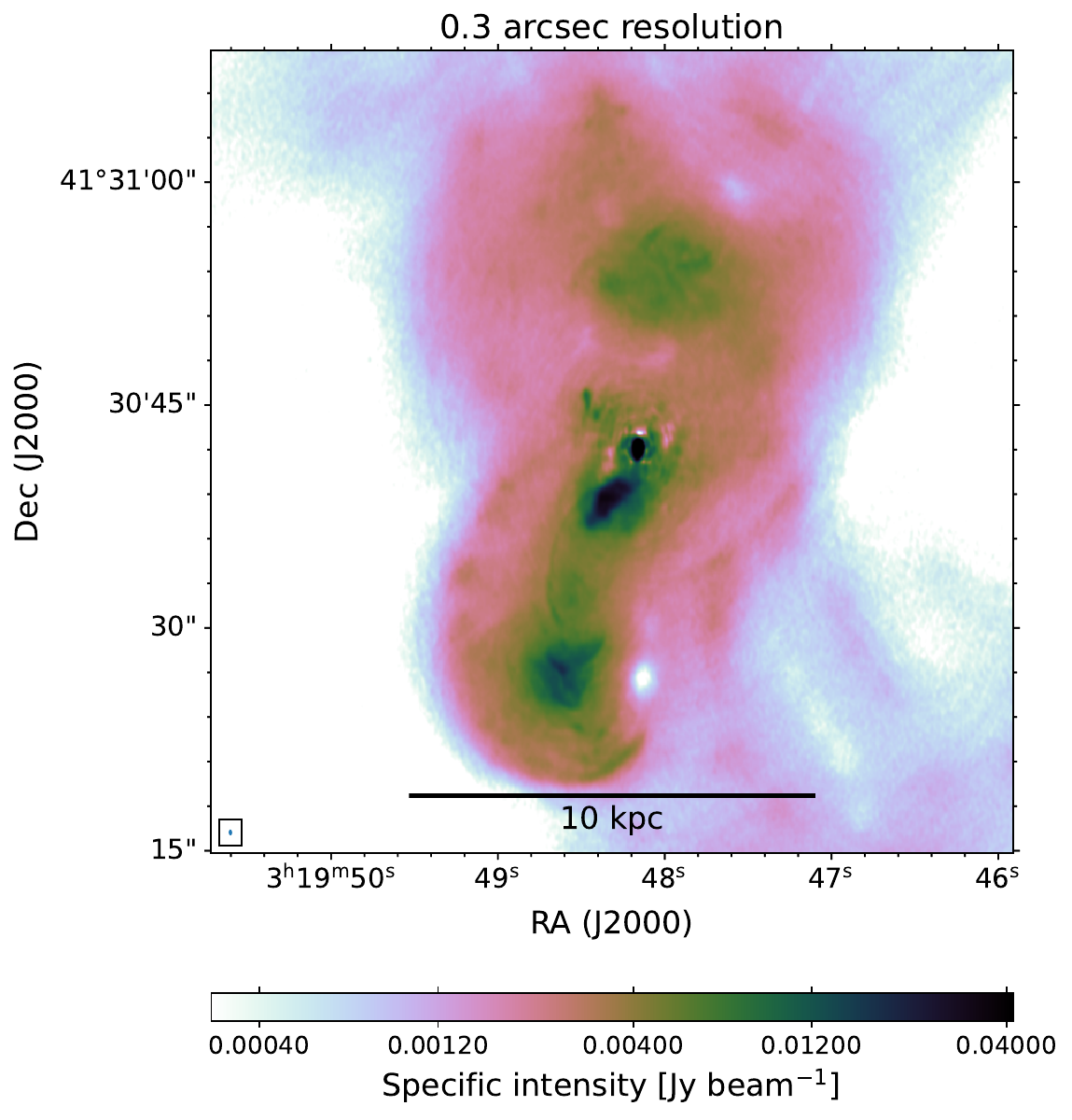}
\caption{High-resolution view of 3C\,84 using all European baselines. Left panel: LOFAR 0.3\arcsec{} resolution image. The inner and outer lobe pairs are labeled. Right panel: Zoomed-in view of the left panel showing the inner structure in more detail. The beam sizes are indicated in the bottom left corners.} 
\label{fig:ILT}
\end{figure*}

\begin{figure*}[t!]
\centering
\includegraphics[width=0.49\textwidth]{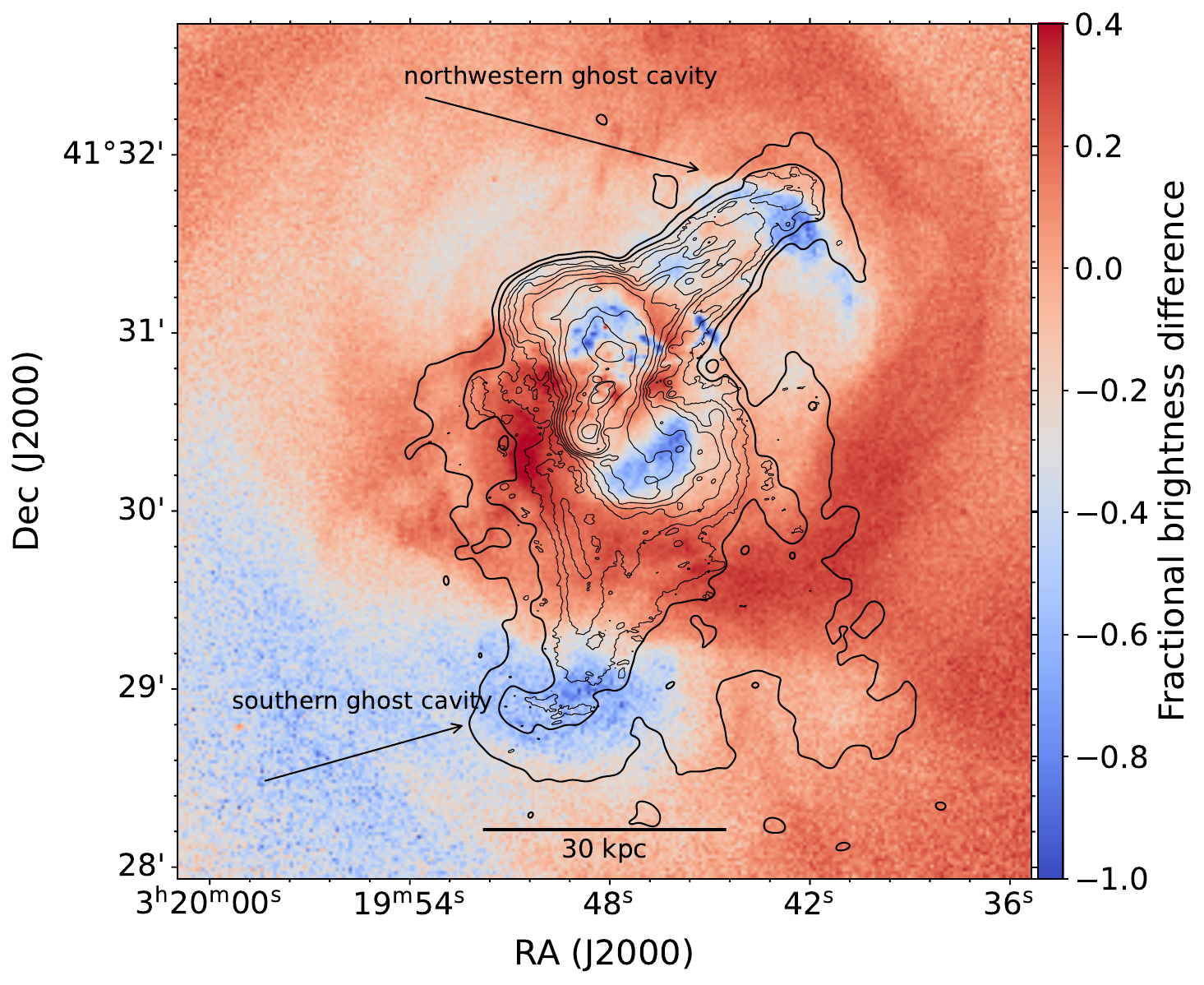}
\includegraphics[width=0.49\textwidth]{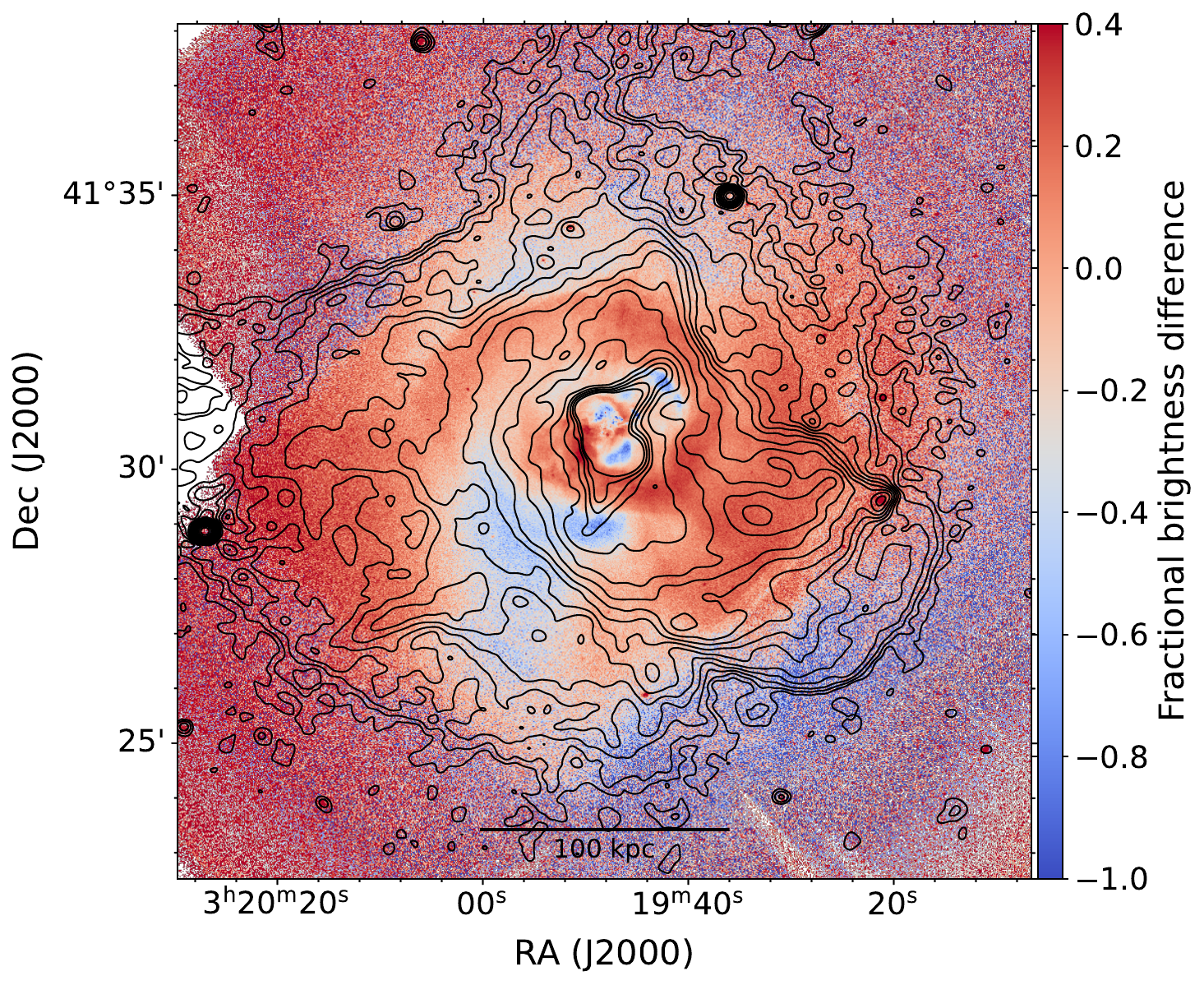}
\caption{\textit{Chandra} 0.5--2.0\,keV fractional residual X-ray images with radio contours overlaid. These images were made by subtracting a radially averaged profile from the original X-ray image and then dividing it by the original. The contours in the left panel are from the 1.6\arcsec{} and 3.4\arcsec{} resolution LOFAR images and drawn at ten logarithmically spaced levels between $10\sigma_{\rm{rms}}$ and $1000\sigma_{\rm{rms}}$ and 
levels of $12\sigma_{\rm{rms}}\times[1,2]$, respectively. The two X-ray ghost cavities are labeled. The contours in the right panel are from the 7\arcsec{} resolution image and drawn at levels of $3\sigma_{\rm{rms}}\times \sqrt{[1,2,4,8,\ldots]}$.}
\label{fig:Chandra_highres}
\end{figure*}

\begin{figure}[t!]
\centering
\includegraphics[width=0.49\textwidth]{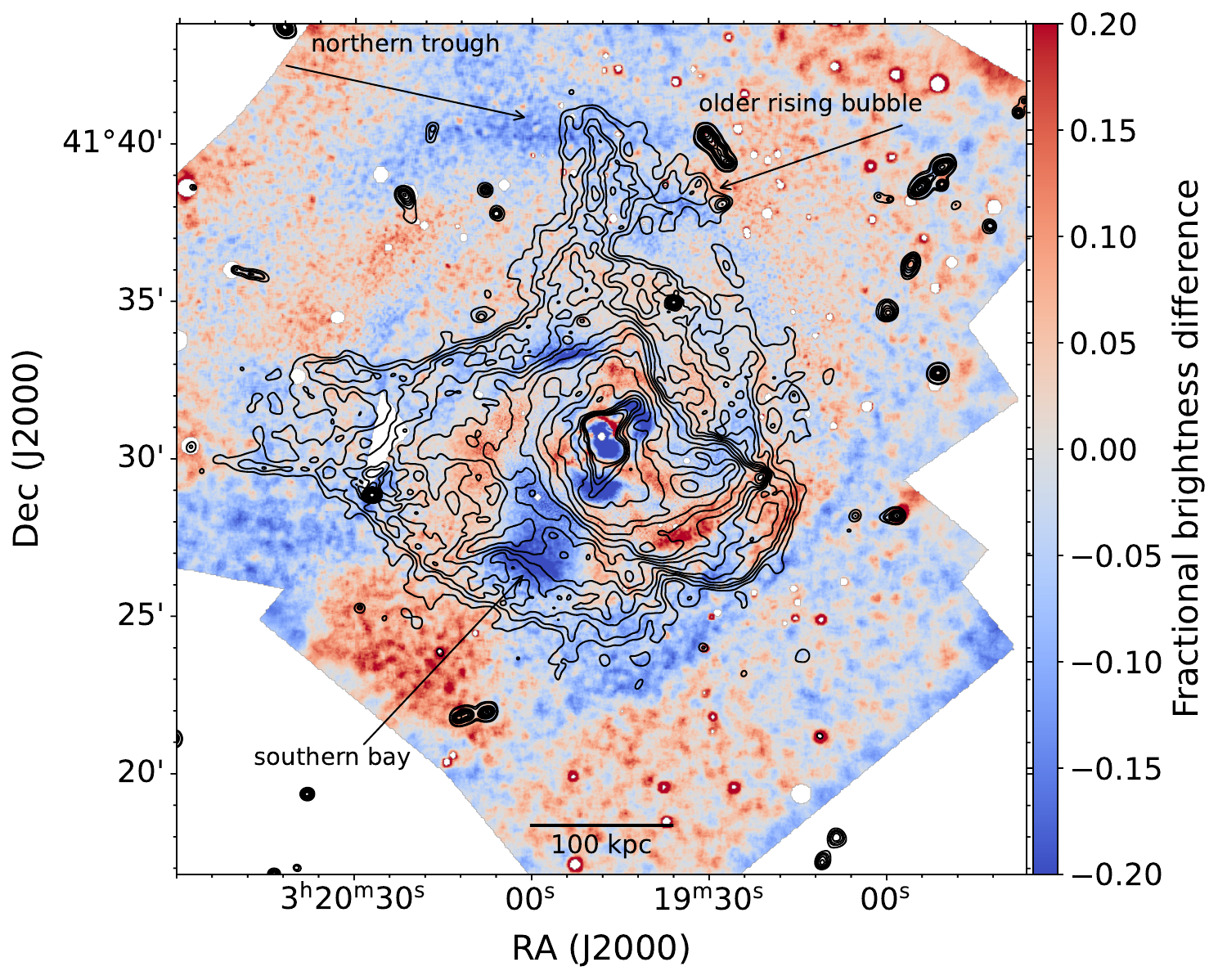}
\caption{\textit{Chandra} fractional residual X-ray image from \cite{2011MNRAS.418.2154F} with 7\arcsec{} resolution LOFAR radio contours overlaid. The contours are drawn at levels of $3\sigma_{\rm{rms}}\times {[1,2,4,8,\ldots]}$.}
\label{fig:Fabian2011}
\end{figure}

\begin{figure*}[t!]
\centering
\includegraphics[width=0.46\textwidth]{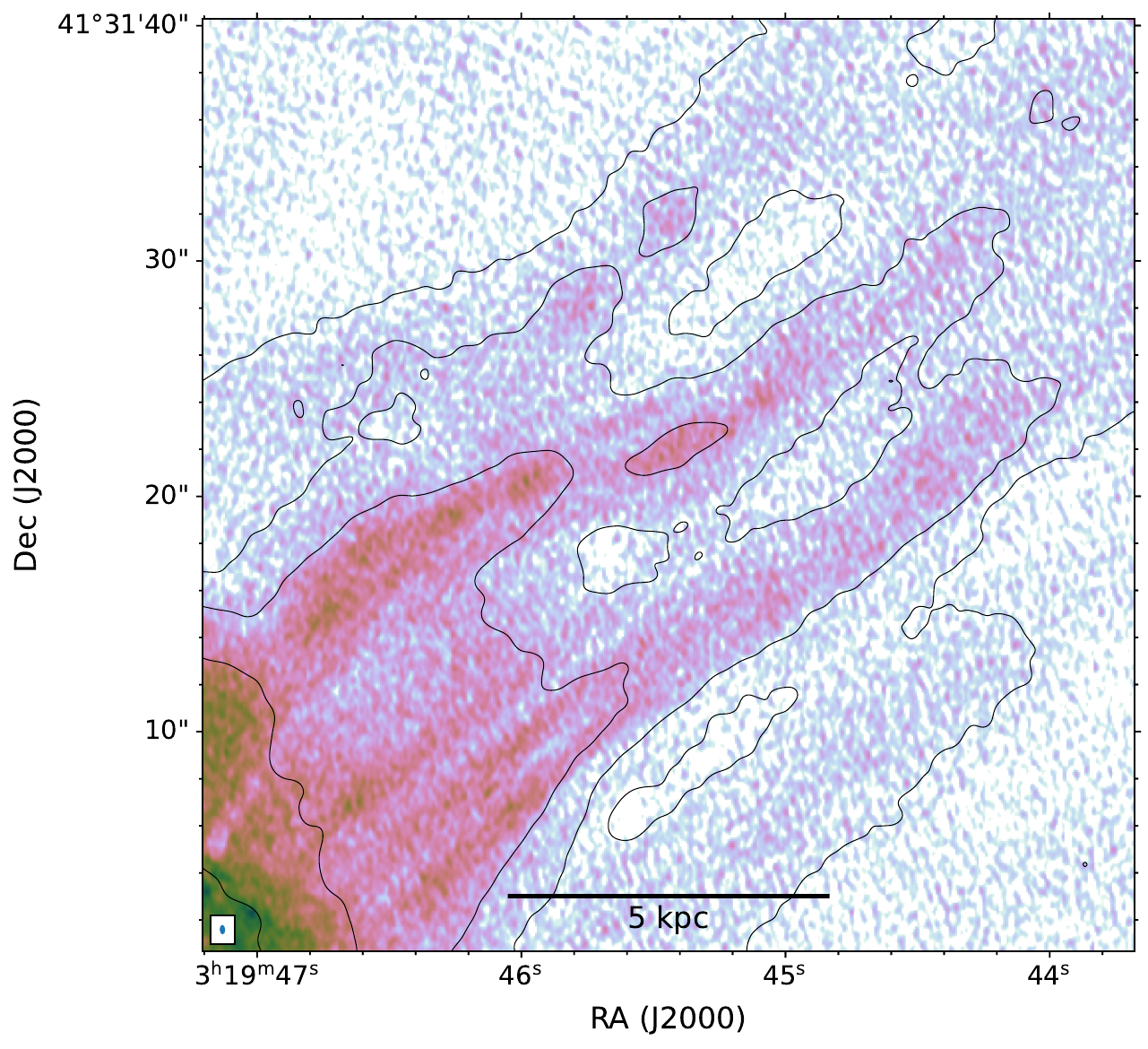}
\includegraphics[width=0.46\textwidth]{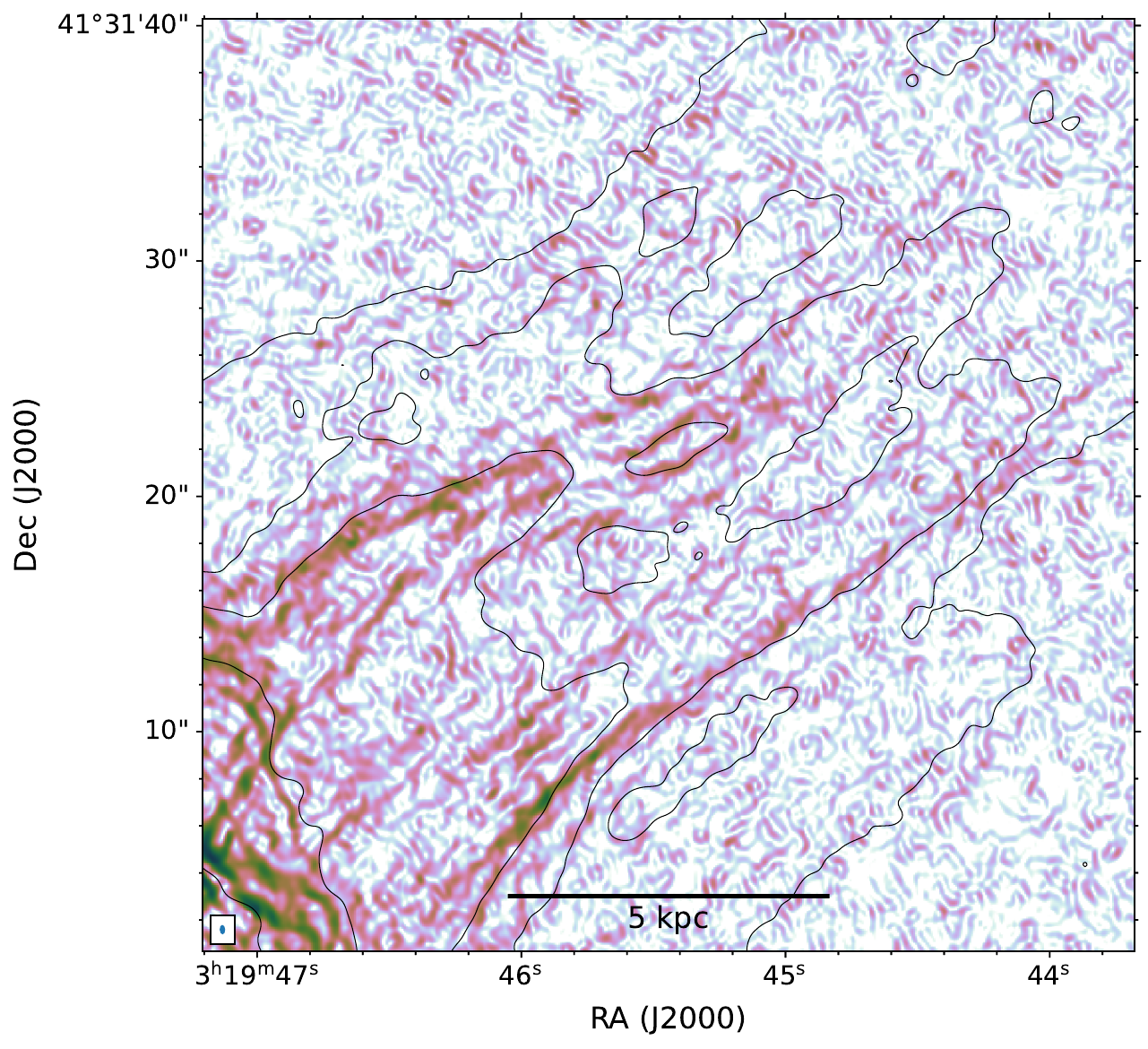}
\caption{144\,MHz 0.3\arcsec{} resolution radio image centered on the extension of 3C84's northern lobe. The left panel image has been processed using unsharp masking with a scale of 1.125\arcsec{} \citep[radius=15 pixels, amount=2;][]{scikit-image}. This technique enhances high-frequency details in the image by subtracting a Gaussian-smoothed version of the original image. Contours from the 1.6\arcsec{} resolution image are overlaid to indicate the location of the four main wider filaments. The contour levels are drawn at $[1,2,4,8,16]\times 0.1$~mJy~beam$^{-1}$ using an additional $5\times5$ pixel kernel smoothing. The right panel image was processed with a GGM filter with $\sigma=0.1875$\arcsec{}. The black contours are the same as in the left panel.
}
\label{fig:unsharp}
\end{figure*}

\begin{figure}[t!]
\centering
\includegraphics[width=0.49\textwidth]{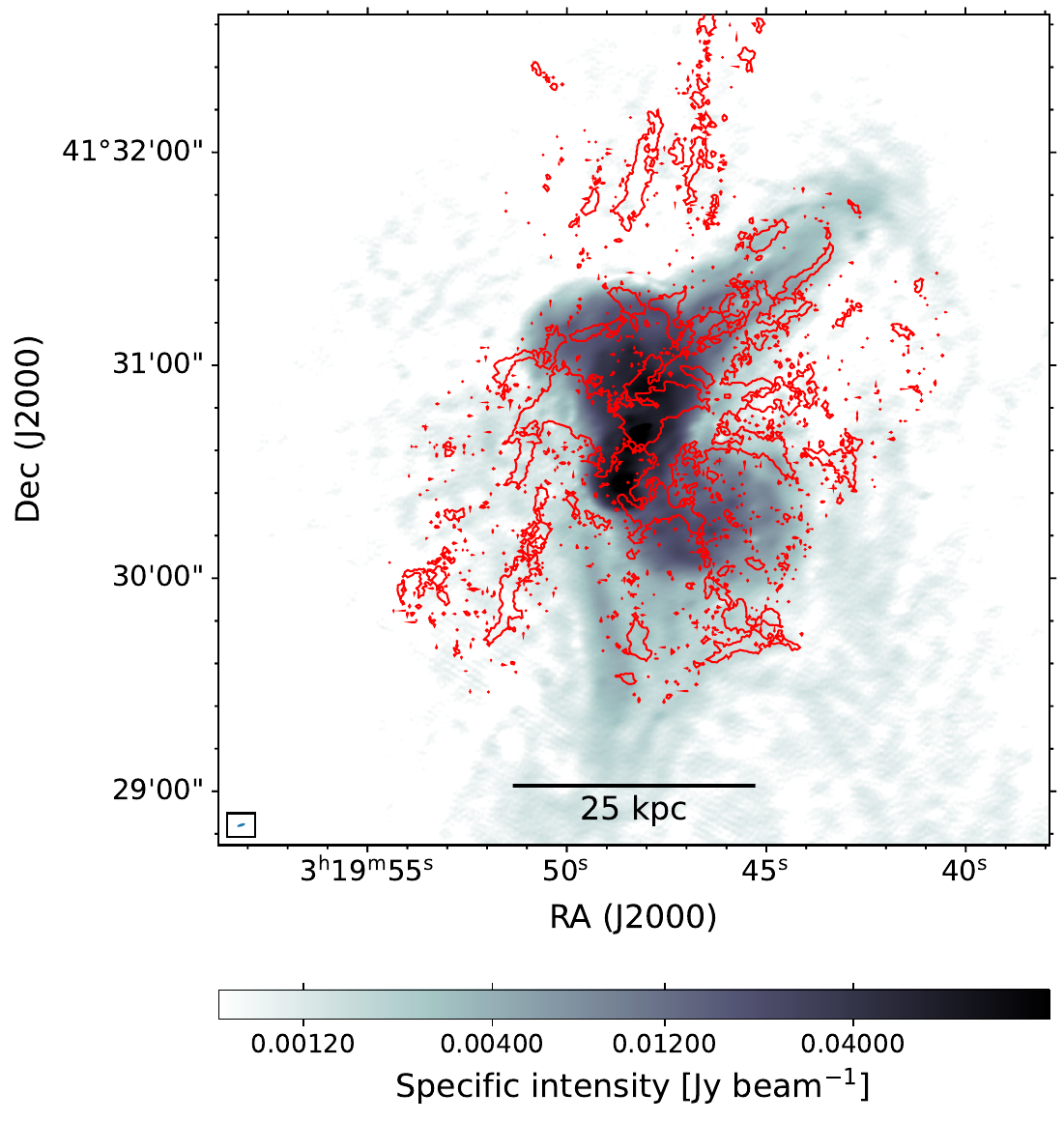}
\caption{144 MHz image of 3C\,84 at 1.6\arcsec{} resolution. The beam size is indicated in the bottom left corner. H$\alpha$ emission line contours are overlaid in red from \cite{2018MNRAS.479L..28G}.}
\label{fig:Halpha}
\end{figure}

\subsubsection{3C\,84 -- NGC\,1275}
The 144\,MHz 0.3\arcsec{} high-resolution image of 3C\,84 reveals a  complex-shaped source (see Fig.~\ref{fig:ILT}; for an additional GGM filtered image, see Appendix~\ref{sec:appendixGGM}). An overlay of the 1.6\arcsec{}, 3.4\arcsec{}, and 7\arcsec{} resolution LOFAR images on the fractional difference X-ray image from the \textit{Chandra} satellite is displayed in Fig.~\ref{fig:Chandra_highres}. The radio source 3C\,84 can be divided into several components (labeled in Figs.~\ref{fig:ILT} and \ref{fig:Chandra_highres}): an inner lobe pair, an outer lobe pair, and radio emission associated with the X-ray ghost cavities, including the radio ``filaments'' leading to the ghost cavities and the radio emission filling the ghost cavities. The filaments to the south and the northwestern ghost cavities have lengths of about 30\,kpc.
In the southern main lobe, we detect loop-like structures, which were also noticed by \citet{2020MNRAS.499.5791G}. The outer lobes of 3C\,84 cover the main pair of X-ray cavities. The various radio components can also (partly) be recognized in the 144--352\,MHz spectral index map (see Fig.~\ref{fig:spix}, right panel). The inner lobe pair has a spectral index of about $-0.5$ (orange color), the outer lobe pair has a spectral index of $-1.0$ (green-blue), and the filaments associated with the outer ghost cavities have spectral indices of $-1.5$ to $-2.0$ (purple). We note that the typical uncertainty for these spectral index values is about 0.15 (see Fig.~\ref{fig:spixerr}, bottom panel).

To study the correspondence between the radio and X-ray emission on a larger scale, we also overlay the 7\arcsec{} resolution LOFAR contours on the fractional difference X-ray image from \cite{2011MNRAS.418.2154F} (see Fig.~\ref{fig:Fabian2011}). In this image, \citeauthor{2011MNRAS.418.2154F} noted the presence of three X-ray depressions in the north, namely two
``outer cavities'' and the ``northern trough''. The trough is possibly the remains of old rising bubbles. A spur of radio emission to the north pointing toward the trough was also noted based on a comparison with WSRT radio data. We also clearly detect this radio emission in our LOFAR image, which we label ``northern extension''. Interestingly, this emission covers the western part of the trough and extends to the west at the location of the northern outer cavity. This lends further support to the scenario where these X-ray depressions are related to old (${\sim}500$\,Myr) radio bubbles. However, given that the radiative lifetime of cosmic-ray electrons is significantly shorter than the estimated ${\sim}500$\,Myr, it is likely that continuous re-acceleration is required to maintain a population of these electrons radiating in the LOFAR band.

The emission leading toward the northwestern ghost cavity consists of at least four individual filaments; the two brighter inner ones were previously detected with the VLA \citep{2020MNRAS.499.5791G}. They have a width of about 1.5--3.0\,kpc. The inner filaments themselves are resolved into several ``fibers'' (see Fig.~\ref{fig:unsharp}). The fibers have a width of $\sim1-1.5$\arcsec{}, corresponding to 350--500\,pc. The detection of several filaments and fibers supports the fact that these structures are likely common for cluster radio galaxies. These structures arise naturally from turbulent MHD flows \citep{2015ApJ...810...93P} and have been observed in other clusters as well \citep[][]{2020A&A...636L...1R,2022ApJ...935..168R}. The origin of the radio-emitting electrons and the steep radio spectrum, however, is not well understood. In this case, we propose a scenario where the filament magnetic fields have been amplified and stretched by the interaction with the radio outflow of 3C\,84.

The central region of the Perseus cluster contains filamentary H$\alpha$ emission line nebulae. These emission line nebulae are also found in other cool-core clusters with peaked X-ray surface brightness
distributions \citep[e.g.,][]{1999MNRAS.306..857C,2012ApJ...746..153M,2019A&A...631A..22O}. Thermal instability is thought to be a key process in the formation of these nebulae \citep{2012MNRAS.419.3319M,2017ApJ...845...80V}. In Fig.~\ref{fig:Halpha}, we display the 1.6\arcsec{} resolution image overlaid with H$\alpha$ emission line contours from \cite{2018MNRAS.479L..28G}. While many of the H$\alpha$ filaments show little relation to the radio emission, we note that some of the  H$\alpha$ filaments seem to straddle the radio emission from the extended southwestern lobe. In addition, several H$\alpha$ filaments are aligned with the radio filaments leading to the northwestern ghost cavity, indicating that there is a connection between the formation of these  H$\alpha$ filaments and the AGN lobes or jets \citep[e.g.,][]{2019A&A...631A..22O}.

\subsubsection{Tailed radio galaxies}

\begin{figure*}[t!]
\centering
\includegraphics[width=0.325\textwidth]{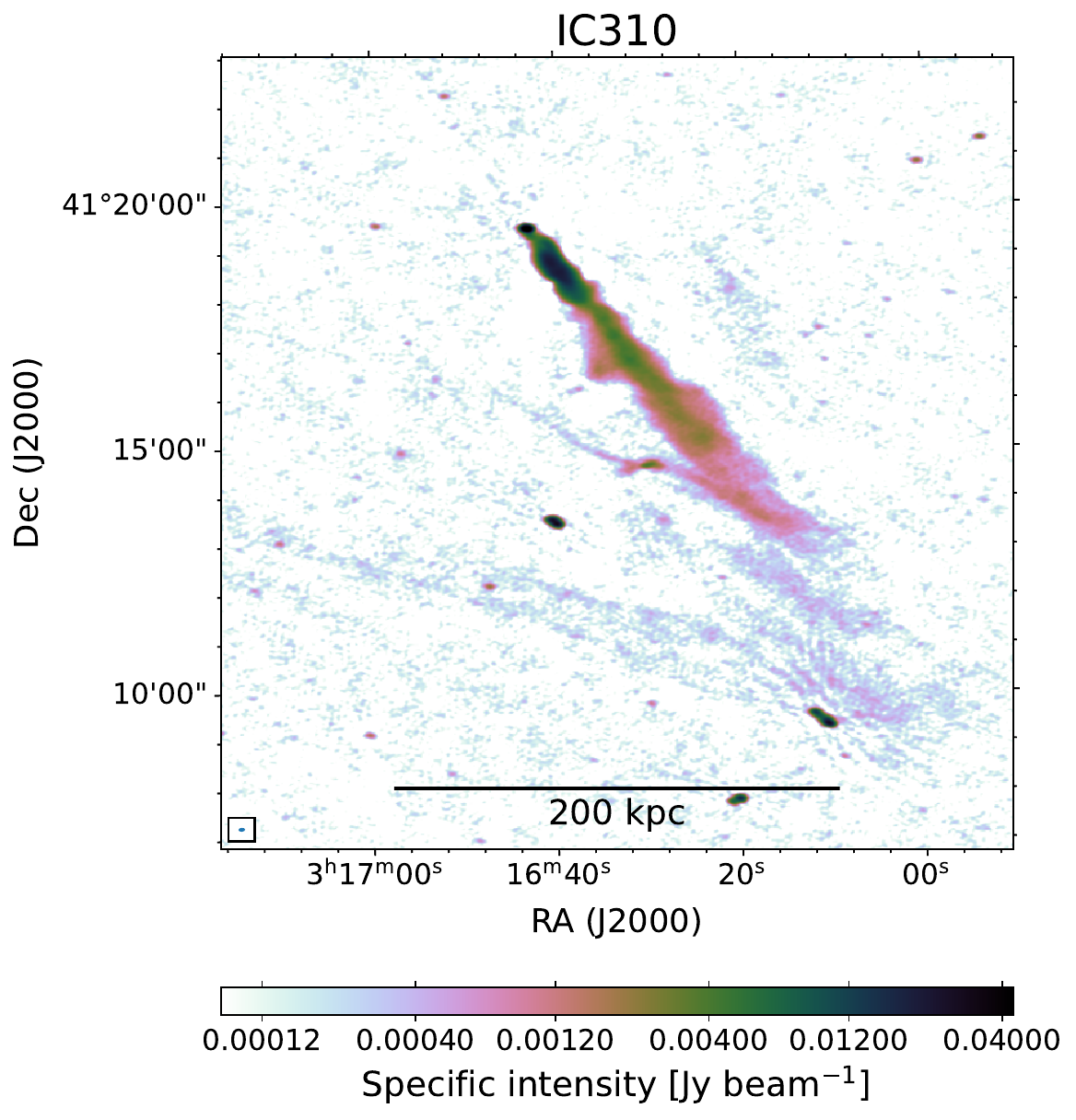}
\includegraphics[width=0.325\textwidth]{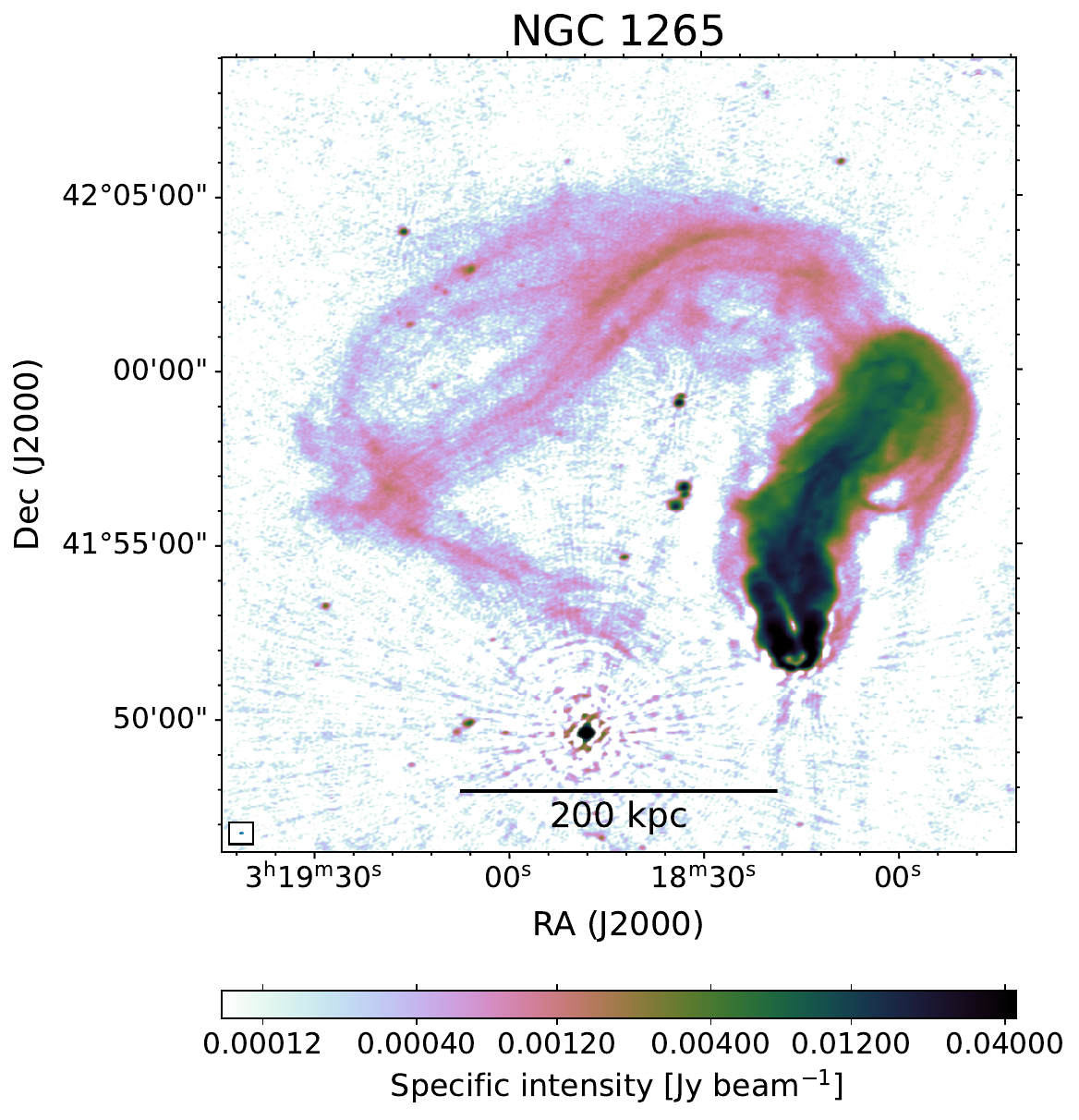}
\includegraphics[width=0.325\textwidth]{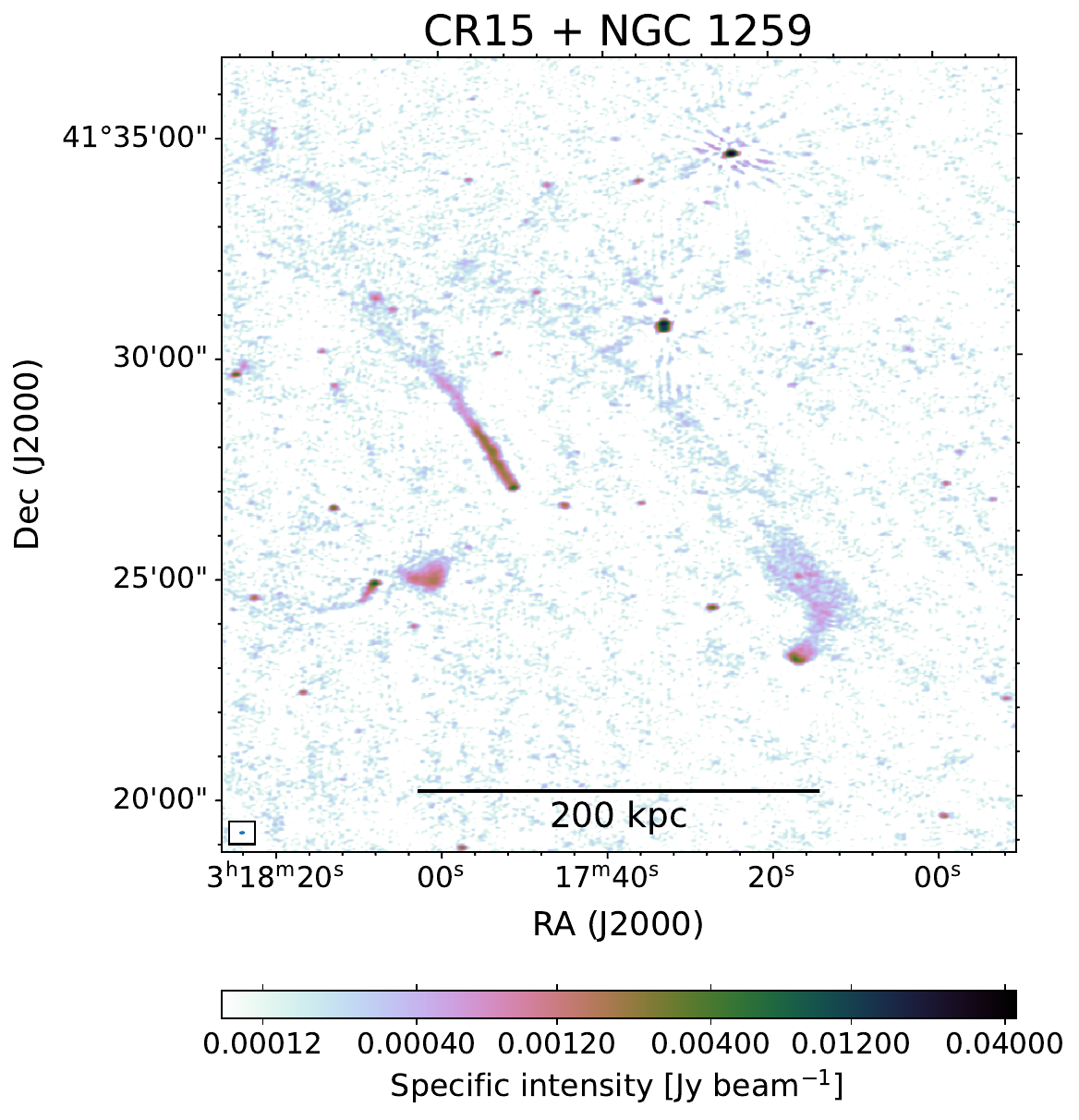}
\caption{144\,MHz radio images around tailed radio galaxies IC\,310 (left),  NGC\,1265 (middle), and CR\,15 + NGC\,1259 (right) at 7\arcsec{} resolution. The tailed sources T\,74, NGC\,1270, and NGC\,1272 are shown in Fig.~\ref{fig:HBAfullzoom}. 
}
\label{fig:tails}
\end{figure*}

\begin{figure*}[t!]
\centering
\includegraphics[width=0.49\textwidth]{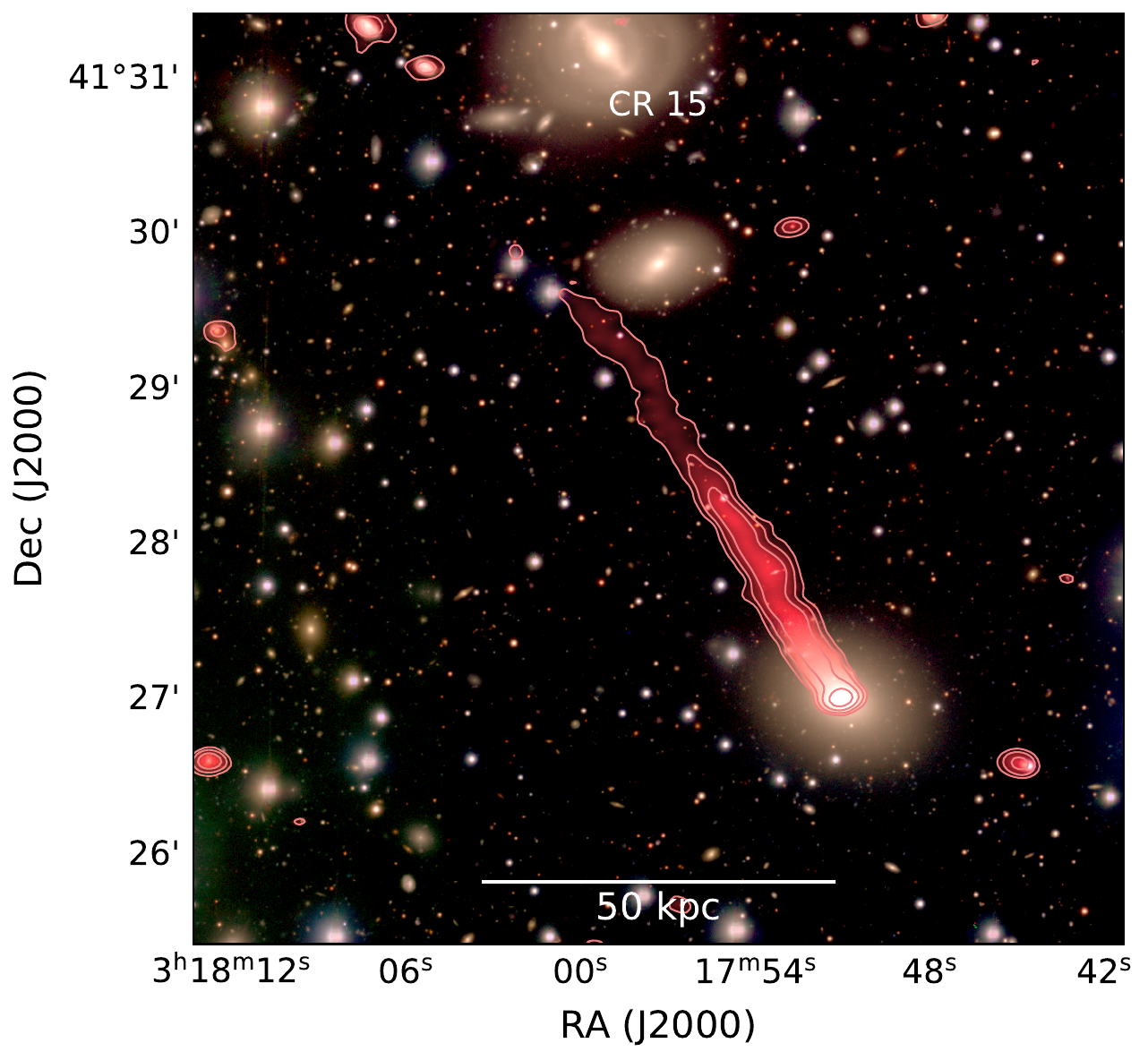}
\includegraphics[width=0.49\textwidth]{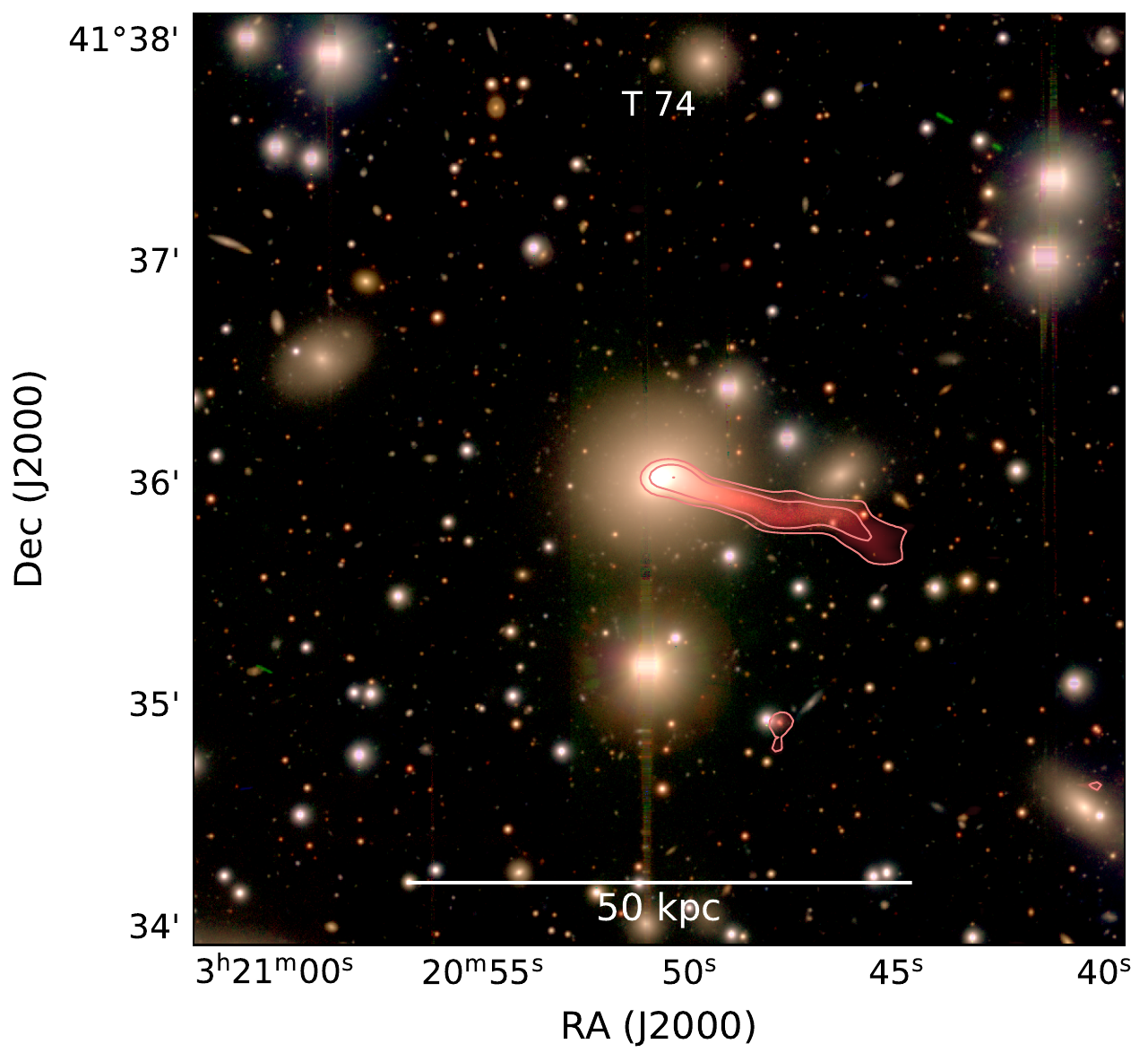}
\includegraphics[width=0.49\textwidth]{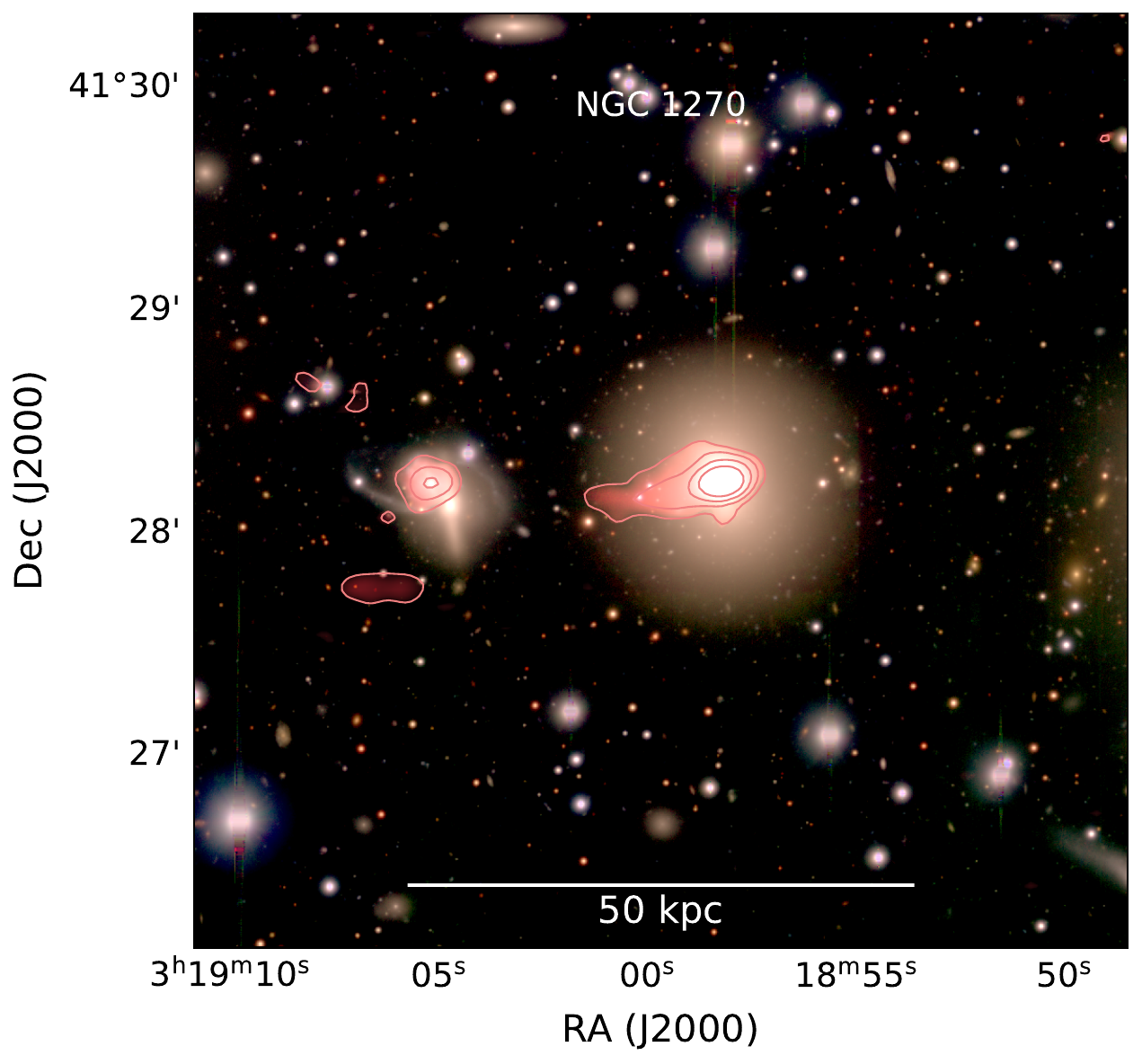}
\includegraphics[width=0.49\textwidth]{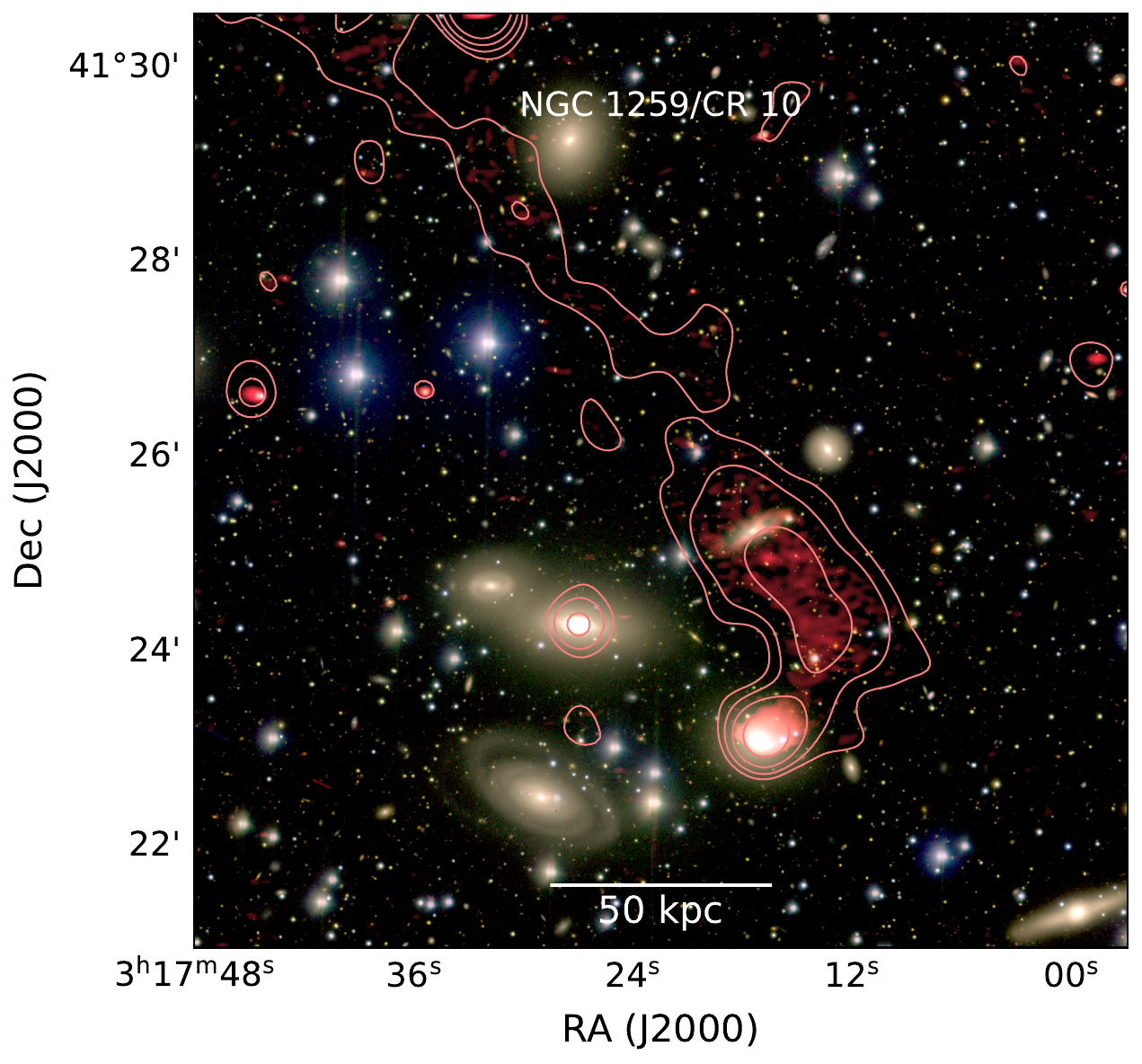}
\caption{Subaru archival $g-$, $r$-, and $i-$ (or $g$, $i$, and $z$ for NGC 1259 as the $r$ -band image had artifacts) band color images obtained from the Hyper Suprime-Cam (HSC) Legacy
Archive \citep{2021PASJ...73..735T}. Radio emission is overlaid in red from the 7\arcsec{} resolution LOFAR image. Red contours are drawn at levels of $[1,2,4,8] \times 4\sigma_{\rm{rms}}$ and are from the same image, except for NGC\,1259, for which we used the 26\arcsec{} resolution image. }
\label{fig:opticaltails}
\end{figure*}

The Perseus cluster is known to host 
several radio galaxies with tailed morphologies. In addition, there are four star-forming member galaxies that display synchrotron tails. The discovery of these so-called jellyfish galaxies was presented in \cite{2022A&A...658A..44R}. Here, we focus on the tailed sources associated with AGNs. In our 7\arcsec{} 144\,MHz image, we detect a total of seven tailed radio AGNs. Among them are the well-known objects NGC\,1265 and IC\,310. 
Cutout images of the tailed sources at 7\arcsec{} resolution are shown in Fig.~\ref{fig:tails}. The sources T\,74, NGC\,1272, and NGC\,1270 are visible in Fig.\ref{fig:HBAfullzoom}. Combined optical and radio overlay images for select sources are shown in Fig.~\ref{fig:opticaltails}. 

Near the western edge of the mini-halo, we observe the tailed AGN NGC\,1272, which was discovered by \cite{2014MNRAS.442..838M} and studied in more detail in \cite{2021ApJ...911...56G}. The tail of this source is blended with the emission of the mini-halo. The tailed source T\,74, discovered by 
\cite{1998A&A...331..901S}, is located to the east of the mini-halo (Fig.~\ref{fig:HBAfullzoom}). Here, we adopt the labeling scheme from \citeauthor{1998A&A...331..901S},  which refers to \cite{1978ApJ...222...54T} for host-galaxy identification. The host galaxy is also known as LEDA\,12547. The 24\,kpc single tail of T\,74 points radially away from the cluster center. At a resolution of 7\arcsec{}, the tail remains unresolved. 
We discover a new tailed radio source associated with the elliptical \object{NGC\,1270} to the west of the cluster center (Fig.~\ref{fig:HBAfullzoom}). The tail of the source points to the east and has a short length of about 14\,kpc. The radio core of this object was previously detected by \cite{1993PhDT.......392S} and \cite{2017MNRAS.465.3943P}.

Tailed sources CR\,15 and \object{NGC\,1259} (also known as  CR\,10) are situated near each other to the west of the cluster center (see Figs~\ref{fig:HBAlowres} and~\ref{fig:tails}). CR\,15 was first detected by \cite{1972Natur.237..269M} and {NGC\,1259} by \cite{1998A&A...331..901S}, with CR referring to \cite{1971ApJ...168..321C}. CR\,15, associated with the galaxy \object{LEDA\,12254},  displays a well-defined 130\,kpc long tail. From the nucleus, the tail of NGC\,1259 first extends about 20\,kpc to the northwest before bending back to the northeast (i.e., the same direction as the tail of CR\,15). The tail of  NGC\,1259 is about 250\,kpc long. In the 26\arcsec{} resolution image, both tails fade and blend into the emission of the giant radio halo (Fig.~\ref{fig:HBAlowres}).

Our LOFAR images also prominently show the tailed AGN  IC\,310. This source has a well-known bright 260\,kpc tail that extends to the southwest (see Fig.~\ref{fig:tails}). The start of the tail near the nucleus of IC\,310 was resolved into two jets by \cite{2020MNRAS.499.5791G}. The 7\arcsec{} and 26\arcsec{} resolution images reveal several filaments in the southern part of the tail. What is most remarkable is that the tail is much more extended than the previous studies showed; it changes direction by about 135\degr{} (this is a projected angle) toward the cluster center. This new ``trail-like'' extension is 700\,kpc long and blends into the emission of the giant radio halo at a cluster-centric distance of $\approx 200$\,kpc. 
The first 200\,kpc of the trail, after the abrupt bending, shows some filamentary substructure. The trail has a relatively constant brightness along its length and is about 200\,kpc wide.

For the narrow-angle tailed source NGC\,1265,  we observe the same general structures as previously detected by, for example, \cite{1986ApJ...301..841O,1998A&A...331..901S,2017MNRAS.469.3872G,
2020MNRAS.499.5791G}. The source consists of two main parts: a bright part with the main lobes and jets originating from the host galaxy and a larger fainter part where the tail bends and loops backward. The reason for this bimodal surface-brightness distribution is discussed in \cite{2011ApJ...730...22P}.
Most notable in the LOFAR image are the many thin filaments visible in the backward bend lobe and the loop-like structure near the top of the main bright lobe. These features are also clearly visible in the GGM-filtered image shown in Appendix~\ref{sec:appendixGGM}.
In Appendix~\ref{sec:backgroundcluster}, we briefly discuss the two tailed radio galaxies that are likely associated with WHL\,J031807.9+412455 \citep[$z_{\rm{phot}}=0.2125$;][]{2012ApJS..199...34W}, a galaxy cluster located behind the Perseus cluster.

\section{Discussion}
\label{sec:discussion}
\subsection{Radio mini-halo}
The LOFAR image of the mini-halo reinforces the findings from \cite{2017MNRAS.469.3872G}, which pointed out that the Perseus cluster mini-halo contains a rich amount of substructure, including filaments and edges. We suggest that this is likely a general feature of mini-halos, but that the lack of deep high-resolution imaging has prevented these substructures from being noticed more often. Two other examples of mini-halos showing some additional substructure are the ones in \object{MS\,1455.0+2232} and \object{RX\,J1720.1+2638}. For \object{MS\,1455.0+2232}, MeerKAT and LOFAR observations by \cite{2022MNRAS.512.4210R} revealed that this mini-halo has an overall asymmetric brightness distribution. Similarly, LOFAR observations of the mini-halo in \object{RX\,J1720.1+2638} showed a more complex brightness distribution, including a possible connection with a tailed radio galaxy \citep{2014ApJ...795...73G,2019A&A...622A..24S,2021MNRAS.508.3995B}. In this case, the radio tail of this galaxy seems to blend with the outer eastern part of the mini-halo. 
For Perseus, the embedded head-tail galaxy NGC\,1272 serves as another indication that radio galaxies, and not only the one associated with the BCG, can affect the morphology of mini-halos. Although, it is hard to completely rule out the possibility of chance projection effects. High-resolution observations of other mini-halos will be needed to assess how often such situations occur. For the Perseus cluster, future polarization observations probing the Faraday depth and dispersion of NGC\,1272 could help determine whether the source is located deep within the ICM or in the foreground \citep[e.g.,][]{2011A&A...525A.104P,2022A&A...657A...2R}. 

It is worth noting that in the case of giant radio halos, recent MeerKAT observations show that edge features are quite common \citep{2023A&A...674A..53B}, with one of the best examples being the Bullet Cluster. In this cluster, the radio edges spatially coincide with edges detected in the X-rays. For the Perseus cluster, the ``eastern spur S'' traces the northern boundary of the concave ``southern bay'' structure seen in X-rays. This X-ray feature could be the result of
a Kelvin–Helmholtz instability, as shown by \cite{2017MNRAS.468.2506W} based on simulations.  Contrary to what was reported by \cite{2017MNRAS.469.3872G},  radio emission is also detected at the location of this bay (see Fig.~\ref{fig:Fabian2011}) thanks to the deeper images.  Edges~1, 2, and 3 lack deep \textit{Chandra} coverage, and no obvious corresponding X-ray features are seen in our image at their precise location; although, the edges' orientations align with the sloshing cold fronts in this cluster  \citep[e.g.,][]{2011MNRAS.418.2154F}. Therefore, part of the complex radio morphology of the mini-halo could be the result of the interaction between the rising lobes of relativistic plasma and the larger-scale sloshing motions and associated turbulence in the cluster core \citep{2022MNRAS.510.4000F}. In contrast, the radio and X-ray edges in the Bullet Cluster are likely all related to a cluster merger event and  not sloshing motions or rising radio bubbles from AGNs.

\subsection{Giant radio halo}
\begin{figure}[t!]
\centering
\includegraphics[width=0.5\textwidth]{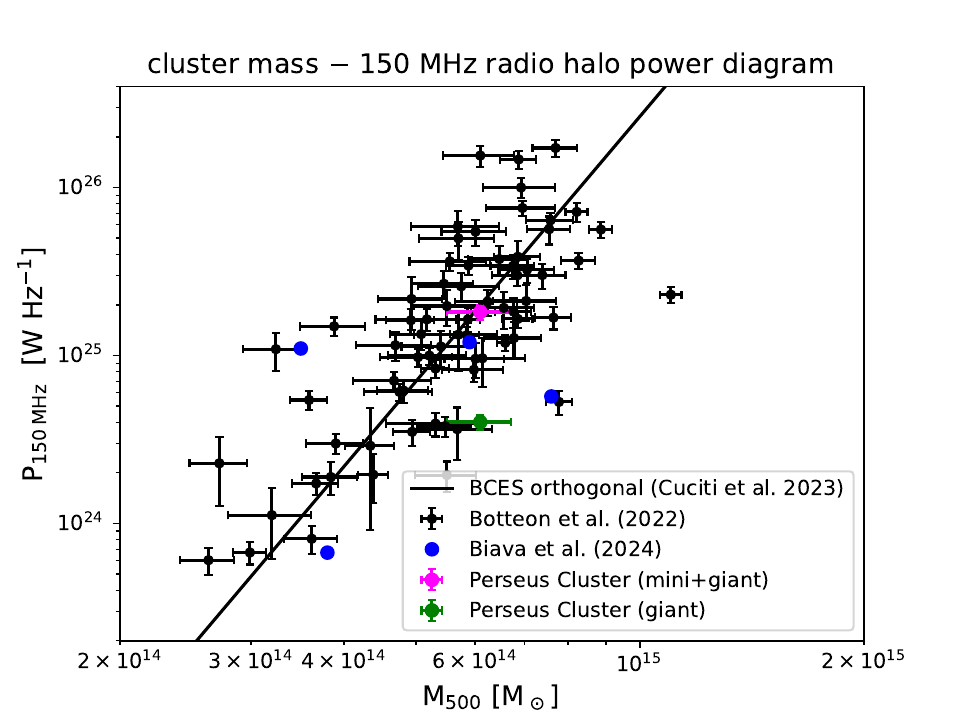}
\caption{150\,MHz radio halo power against cluster mass (M$_{500}$). The black data points were taken from \cite{2022A&A...660A..78B}. The blue points are from \cite{2024A&A...686A..82B}, where the emission from the mini-halo component was removed.
The solid line shows the best-fit correlation using the BCES orthogonal method from \cite{2023A&A...672A..43C}. The magenta data point is for the combined emission from the radio halo and the mini-halo. The green point is for the giant radio halo only. The flux densities to compute the radio power came from our radial surface-brightness profile fitting (Eq.~\ref{eq:profile}), integrated out to $3r_{e}$. For the blue data point, we simply added the individual ``halo components'' integrated out to their own specific $3r_{e}$ values.}
\label{fig:MPdiagram}
\end{figure}

The discovery of diffuse radio emission with a projected size of 1.1\,Mpc indicates the entire Perseus cluster is filled with cosmic rays. The central surface brightness  of $I_{\rm{0}} = 6.3\pm0.7$\,$\mu$Jy\,arcsec$^{-2}$ and characteristic scale of $r_{\rm{e,outer}}=156\pm1.6$\,kpc are very similar to other giant radio halos detected by LOFAR at 144\,MHz \citep{2022A&A...660A..78B}. For giant radio halos, there is a well-known correlation between cluster mass (and its various proxies) and radio power \citep[e.g.,][]{2000ApJ...544..686L,2013ApJ...777..141C,2015A&A...579A..92K,2012MNRAS.421L.112B}. The dependence of the radio power on cluster mass should reflect that a fraction of the gravitational energy released during cluster merger events is channeled into the re-acceleration of cosmic rays by ICM turbulence \citep[e.g.,][]{2005MNRAS.357.1313C,2023A&A...672A..43C}.  The current best measurement of this $\rm{M}_{500} -P_{\rm{150~MHz}}$
correlation for a mass-selected sample of clusters at 150\,MHz comes from the study by 
\cite{2023A&A...672A..43C}. If we put the giant radio halo component of the Perseus cluster on a $\rm{M}_{500} -P_{\rm{150~MHz}}$ diagram, using the X-ray-derived cluster mass from \cite{2017ApJ...841...71G} we find that it falls slightly below the best-fit correlation (see Fig.~\ref{fig:MPdiagram}). %
If we also include the  mini-halo radio power, this moves the radio power on top of the correlation. Irrespective of the inclusion of the radio mini-halo, the radio power seems to be consistent with the mass-power relation given the intrinsic scatter of this relation. Thus, both in terms of characteristic size and radio power, the Perseus cluster halo has similar properties to other known giant radio halos observed by LOFAR in the 120--168\,MHz frequency band. 

Part of the scatter in the $\rm{M}_{500} -P_{\rm{150~MHz}}$ correlation arises from measurement errors in mass and radio power. However, beyond this measurement scatter, the correlation likely has intrinsic scatter due to inherent variations caused by clusters being in different dynamical stages, with some mergers injecting more turbulence than others, and radio halos having different spectra and sizes \cite[e.g.,][]{2023A&A...680A..30C}. Additionally, clusters evolve between merging and more relaxed states, with synchrotron emission being amplified during mergers and suppressed as clusters become more relaxed \citep{2009A&A...507..661B}. The effect of a cluster’s dynamical state on its location above or below the correlation was investigated by \cite{2023A&A...680A..30C}, who found that radio halos located above the correlation are indeed typically found in more dynamically disturbed clusters, and vice versa. In the case of Perseus, the fact that its radio power falls below the correlation could indicate that, in a global sense, the cluster is dynamically less disturbed than those that fall on or above the correlation. This interpretation is consistent with the fact that the cluster shows evidence of an off-axis merger event that left the cool core intact (see Sect.~\ref{sec:originturbulence}).

Recently, \cite{2024A&A...686A..82B} investigated the presence of diffuse radio emission beyond cluster cores in a sample of relaxed clusters. In about 30\% of these clusters, radio emission was detected beyond the cluster core, with the corresponding radio power being consistent with the $\rm{M}_{500} -P_{\rm{150~MHz}}$ relation, similar to Perseus. For the non-detections, it was concluded that the derived upper limits were not sufficiently deep to rule out their presence within the expected scatter of the correlation. Considering that the Perseus cluster is the nearest example of a cool-core system, recent discoveries of radio emission beyond the classical extent of mini-halos \citep{2014MNRAS.444L..44B,2017A&A...603A.125V,2018MNRAS.478.2234S,2019A&A...622A..24S,2021MNRAS.508.3995B,2022A&A...657A..56K,2023A&A...678A.133B,2024A&A...683A.132L,2024A&A...686A..82B} implies that giant radio halos are relatively common (maybe even ubiquitous) in massive cool-core clusters and not only merging systems. However, uncovering (most of) these giant radio halos would require high--dynamic range imaging (to reduce contamination from the bright central AGN), and preferentially low-frequency observations. That said, despite the presence of a cool core in the Perseus cluster, there is also evidence of dynamical activity. Therefore, the simple merger versus relaxed classification in the context of giant radio halos might be an oversimplification (see the next section).

\subsubsection{Origin of the turbulence}
\label{sec:originturbulence}
The presence of cluster-scale diffuse emission in the Perseus cluster has important implications for the X-Ray Imaging and Spectroscopy Mission (XRISM) mission \citep{2020arXiv200304962X,2016Natur.535..117H}. In light of the turbulent re-acceleration model, this indicates that there is sufficient turbulence present in the cluster outskirts for the re-acceleration to happen. The fact that the radio halo encompasses the entire cluster also indicates that particle re-acceleration is not localized to particular regions. 

A question is if the giant radio halo in the Perseus cluster can be classified as an ``intermediate-phase'' halo. It has been proposed that these intermediate-phase halos are located in clusters that show a moderate amount of dynamical activity and are transitioning between relaxed and merging states \citep[e.g.,][]{2014MNRAS.444L..44B,2017MNRAS.466..996S,2019MNRAS.486L..80K,2020MNRAS.493L..28R}. An important consideration therefore revolves around the origin of the turbulence in the outskirts of the Perseus cluster. In one scenario, turbulence is injected in the cluster volume by continuous minor mergers with small galaxy groups \citep[e.g.,][]{2023A&A...678A.133B}. These mergers then produce a large-scale radio halo with an ultra-steep spectrum by turbulent re-acceleration. In another scenario, the Perseus cluster has undergone a specific merger event in the past and the turbulence injected by this event has not yet completely dissipated. Interestingly, the overall ICM distribution of the Perseus cluster is known to be asymmetric on large scales, and \cite{2012ApJ...757..182S} discovered the presence of sloshing cold fronts outside of the cluster core. These reach a distance of about 1\,Mpc from the cluster center, with an estimated age of about 5\,Gyr \citep{2018NatAs...2..292W} for one of the cold fronts. More recently, \cite{2022ApJ...929...37W} found two X-ray surface-brightness edges at 1.2 and 1.7\,Mpc to the west of the cluster center with XMM-Newton, with Suzaku temperature measurements being consistent with these edges being cold fronts. By comparing to simulations, they found that these cold fronts could have originated from a merger event with a large impact parameter of about 1\,Mpc and a low mass
ratio (${\sim}3$). The core passage of this event should have happened almost 9\,Gyr prior. Most recently, through a weak lensing analysis \cite{2024arXiv240500115H}  concluded that the Perseus cluster underwent a major off-axis merger event (mass ratio 3:1) about 5.5\,Gyr ago, further supporting the fact the Perseus cluster is not fully relaxed. Therefore, the sloshing motions and turbulence induced by a past merger event may still generate enough turbulence in Perseus to power a giant radio halo with an ultra-steep spectrum.

Given that the proposed merger events in the Perseus cluster happened a long time ago ($\gtrsim5$\,Gyr ago), it is likely that such large sloshing cold fronts in cluster outskirts are quite common in ``relaxed'' clusters \citep[see also][]{2013A&A...556A..44R,2014MNRAS.441L..31W,2023MNRAS.526L.124M}, suggesting that large radio halos should also be commonly found at low frequencies in such objects \citep[see, e.g., the recent analysis of Abell\,2142][]{2024A&A...686A..44R}.

It is important to determine the spectral index of the giant radio halo in Perseus. 
Studies indicate that the diffuse emission beyond the cores of (semi-)relaxed clusters has a steeper spectral index \citep[e.g.,][]{2017A&A...603A.125V,2018MNRAS.478.2234S,2021MNRAS.508.3995B,2023A&A...678A.133B,2024A&A...686A..44R} than radio mini-halos. However, some of the observed clusters only show extended emission beyond the cluster core in particular directions (and are not fully ``enveloping'' giant radio components at the depths studied). A steeper spectral index beyond the cool-core region would be a natural consequence of the turbulent re-acceleration model if there is less turbulent energy available compared to the sloshing core
in these systems \citep[e.g.,][]{2005MNRAS.357.1313C,2008Natur.455..944B}. Similarly, it is also expected that the level of turbulent energy beyond these cores is less than those in merging clusters. Therefore, giant radio halos in cool-core clusters might be on average steeper than their cousins in merging systems. This is also in line with the finding that the fraction of LOFAR-discovered radio halos is larger in less disturbed systems \citep{2023A&A...672A..43C}, which would be the case if these halos have steeper spectra and are thus more easily detected at lower frequencies.

\subsubsection{Halo--tail connection}
Turbulent re-acceleration requires the presence of seed electrons that are already relativistic \citep[see the review by][]{2014IJMPD..2330007B}. In this light, the discovery of a long trail of emission behind IC\,310 blending into the emission from the radio halo indicates AGNs are a natural source for seed cosmic-ray electrons. The overall path of the IC\,310 tail and trail, suggests that the host galaxy is on an elliptical orbit around the cluster, with the galaxy being past the orbit's pericenter and falling back toward the cluster center. On the other hand, abrupt changes in the direction of tails have been observed in other clusters \citep[e.g.,][]{2015MNRAS.452.3064R,2019A&A...622A..25W,2024A&A...686A..44R}, which are still poorly understood and could also be indicative of interactions with the local environment. In the case of IC\,310, the trail has a relatively constant surface brightness along its extent of 700\,kpc. This cannot be directly reconciled with the effects of spectral aging, for which we would expect the trail brightness to decrease. Therefore, the plasma in the trail is likely continuously being re-accelerated, with a balance between losses and re-acceleration being achieved. Re-acceleration has also been observed for some other tailed sources \citep[e.g.,][]{degasperin+17,2024MNRAS.528..141L} and is also supported by simulations \citep[e.g.,][]{2023ApJ...951...76O}. The precise nature of the re-acceleration mechanism is unclear, but this would be an inefficient process for explaining the observed radio properties (e.g., stochastic Fermi II re-acceleration by turbulence). Mapping the spectral index along the trail with future observations is crucial to test the re-acceleration scenario and confirm that there is no gradient due to spectral aging.

Besides IC\,310, the tails of radio sources associated with NGC\,1259 and CR\,10 also blend into the radio halo, locally enhancing its emission. Similarly, 3C\,84 and NGC\,1272 are both fully embedded within the mini-halo structure. Disentangling the emission components in all these areas is, however, challenging. Despite these uncertainties, the observation of five radio galaxies morphologically connected to the giant radio halo and mini-halo, and considering that over a cluster's lifespan it hosted additional radio galaxies, suggests that radio galaxies may be the primary source of seed cosmic-ray electrons in the ICM. This is supported by calculations from \cite{2024Galax..12...19V}. They found that the time-integrated activity of a cluster radio galaxy population appears sufficient to fuel the entire extent of diffuse radio emission with the required amount of fossil electrons.

\subsection{Radio relics}
The existence of the two radio relics to the north and south of the cluster center, in addition to that of the giant radio halo, provides evidence of dynamical activity in the cluster outskirts. The relics in the Perseus cluster are located at similar projected distances to other known relics \citep[e.g.,][]{2012MNRAS.421.1868V,2017MNRAS.470..240N,2023A&A...680A..31J}. Large radio relics trace shock waves in the ICM where particles are (re-)accelerated. Shocks related to major merger events typically have convex shapes as they trace  outward-traveling shock waves \citep[e.g.,][]{
1998A&A...332..395E,1999ApJ...518..603R}. Prominent examples are double relics located along the merger axis, diametrically opposite from the cluster center \citep[e.g.,][]{vanweeren+10,2017ApJ...838..110G,2021MNRAS.505.4762J,2023arXiv230411784K}. There are no clear examples of such relics in relaxed cool-core clusters.

The relics in Perseus, however, do not show clear convex shapes. Instead, the relic RN has a concave shape. A few other relics are known that have concave shapes. These have also been called "wrong-way" or "inverted" relics \citep[e.g.,][]{2020ApJ...900..127H,2021ApJ...914L..29B,2022MNRAS.515.1871R,2023arXiv231106340L}. Radio relics can have more complex shapes as merger events involve multiple smaller substructures \citep[e.g.,][]{2013ApJ...765...21S,2017MNRAS.464.4448W,2021MNRAS.506..396W}. It has been proposed that inverted relics form when an outward-traveling shock wave is bent inward by an infalling galaxy group or cluster \citep{2023ApJ...957L..16B,2023arXiv231106340L}. 

We speculate that the relics in the Perseus cluster have a different origin compared to ``classical''  relics found in highly disturbed merging clusters \citep[e.g.,][]{vanweeren+19}. The Perseus cluster relics might be related to local dynamical activity; for example, smaller merger events or large-scale accretion of matter along a cosmic filament, since there is no evidence for any major recent ($\sim 1$\,Gyr) north-south merger event from X-ray observations. We also note that the relics do not align with the large-scale ICM sloshing patterns as observed by \cite{2012ApJ...757..182S} and \cite{2018NatAs...2..292W}. The location of relic RN is also interesting as it is placed relatively near the head-tail galaxy NGC\,1265 at a similar projected distance to the cluster center. \cite{2011ApJ...730...22P} argued that the unusual morphology of the head-tail radio galaxy can be explained by a past passage of the galaxy through an accretion shock wave. Their proposed shock location agrees well with the location of relic RN, and a possible scenario could be that the accretion shock has re-accelerated regions of fossil radio plasma from another old radio galaxy. Sensitive Suzaku observations by \cite{2021A&A...652A.147Z} do report evidence of a shock in the northwestern outskirts of the cluster. However, the location of this shock is much further, out at a projected radius of 1.8\,Mpc, and it is thus probably unrelated to RN.

More generally, recent deep observations indicate that significant regions of cluster outskirts can be sites of particle (re-)acceleration, where kinetic energy is transferred into relativistic particles and magnetic fields via shocks or turbulence, or both \citep[][]{2022Natur.609..911C,2022SciA....8.7623B}. Both relics, RN and RS, could be the brightest locations where this emission is starting to manifest itself in the Perseus cluster. On the other hand, a scenario where these features are remnants of aged AGN lobes is hard to fully exclude based in the current observations.
Future spectral mapping and polarization observations will thus be crucial to constrain the origin of RN and RS. However, this will be challenging given their very low surface brightness and the presence of 3C\,84 relatively nearby.


\section{Conclusions}
\label{sec:conclusion}
In this paper, we present 120--168\,MHz LOFAR observations of the Perseus cluster. These observations probe a large range of angular scales, corresponding to physical scales of 100\,pc to several megaparsecs. By subtracting the 0.3\arcsec{} resolution model of 3C\,84 made with LOFAR's international baselines, we obtained high-fidelity images (a dynamic range as high as  $8\times10^{4}$) of the extended emission in the cluster center. Below, we list our main findings.

\begin{itemize}
    \item We discovered a giant radio halo in the Perseus cluster. This giant radio halo has distinct properties from the previously known radio mini-halo. The giant radio halo has a projected size of 1.1\,Mpc, with a characteristic e-folding radius of $156\pm2$\,kpc, which is about five times larger than the mini-halo. For this giant radio halo, we compute a 150\,MHz radio power of $(4.0\pm0.5)\times 10^{24}$~W~Hz$^{-1}$. 
    \item The radio mini-halo displays a significant amount of substructure in a 7\arcsec{} resolution image, with an overall asymmetric brightness distribution and several edges being present. In addition, it contains a narrow (3.7 \,kpc) and long (170\,kpc) filament extending to the east.
    \item At projected cluster-centric distances of 0.7 and 1.0 Mpc, two irregular-shaped diffuse sources are detected. The sources have projected lengths of about 400 to 500\,kpc, with the northern source having a concave shape and the southern one consisting of two substructures. Given their peripheral location, projected length, and lack of optical counterparts, we classify these as radio relics. We note, however, that a scenario in which these features are remnants of aged AGN lobes is hard to fully exclude based on the current observations.
    \item We detected the presence of a 700\,kpc trail of radio emission behind the tailed radio galaxy IC\,310. This trail connects to the previously known tail of IC\,310 and extends toward the cluster center, where it blends with the giant radio halo. This trail has a relatively constant brightness along its length and is about 180\,kpc wide. Along its length, the trail displays some filament-like structures. Similarly to IC\,310, we find that the tails of the cluster radio galaxies CR\,15 and NGC\,1259 blend with the giant radio halo. 
    \item We discovered a tailed radio galaxy associated with NGC\,1270. It has a short eastward-pointing tail with a length of 14\,kpc.
    \item Comparing our LOFAR images with \textit{Chandra} X-ray observations, we find that the two X-ray ``ghost'' cavities in the cluster, northwest and south of 3C\,84's nucleus (associated with NGC\,1275), are filled with radio plasma. This radio plasma is connected to 3C\,84 by filament-like structures and has a steep spectrum, with a spectral index of $\alpha=-1.5$ to $-2.0$ measured between 144 and 352\,MHz. On a larger scale, we observe a correspondence between the radio emission and the candidate old radio bubbles in the north identified by \cite{2011MNRAS.418.2154F}.
    \item The steep spectrum radio emission leading to the northern ghost cavity consists of four parallel filaments, with the two brighter inner filaments being resolved into several fibers, or sub-filaments. The origin of these fibers gives rise to new theoretical questions. The fibers have a width of $\sim 1-1.5$\arcsec{}, corresponding to 350–500\,pc. The fibers show spatial alignment with optical H$\alpha$ emission line nebulae \citep[e.g.,][]{2018MNRAS.479L..28G}.

\end{itemize}

Our observations further highlight the complex nature of the Perseus cluster radio mini-halo as revealed by \cite{2017MNRAS.469.3872G}. The Perseus cluster's proximity and it being the mini-halo with the highest quality radio imaging suggest that other mini-halos likely have a significant amount of substructures. The scarcity of additional examples is thus explained by the lack of high-spatial-resolution and signal-to-noise images of other mini-halos. This situation is somewhat similar to that of giant radio halos, where recent MeerKAT images also display a considerable amount of substructures \citep{2023A&A...674A..53B}. Furthermore, the observations of several synchrotron filaments or threads in 3C\,84, NGC\,1265, and other recent detections \citep[e.g.,][]{2022ApJ...935..168R,2020A&A...636L...1R} also indicate that such structures might be common for cluster radio galaxies when deep high-resolution images are available.

The discovery of a giant radio halo in the Perseus cluster indicates that there is sufficient turbulent energy in the cluster outskirts for particle re-acceleration. The origin of this turbulence remains unclear, but it could be related to a past off-axis merger event. Further evidence of dynamical activity in the Perseus cluster outskirts is provided by the discovery of two possible relics to the north and south of the cluster center. Our work also highlights that defining a cluster as either relaxed or merging is overly simplistic when explaining the presence of diffuse radio emission. The specifics of the merger event -- such as whether it was head-on or off-axis -- may result in the coexistence of a radio mini-halo in the cooling core, while a giant radio halo traces the off-axis merger.
An implication from the turbulent re-acceleration model is that giant radio halos in clusters with ``relaxed'' cores should, in general, have steeper spectra than the halos in major mergers since less turbulent energy is available. LOFAR low-band antenna observations will be important to test this prediction.

\begin{acknowledgements}
We thank Andrew Fabian for providing the {Chandra} fractional residual X-ray image and Martijn Oei for comments on the manuscript. RJvW, RT, and CG acknowledge support from the ERC Starting Grant ClusterWeb 804208. RT is grateful for support from the UKRI Future Leaders Fellowship (grant MR/T042842/1). This work was supported by the STFC [grants ST/T000244/1, ST/V002406/1]. ABot and ABon acknowledge financial support from the European Union - Next Generation EU. IDR acknowledges support from the Banting Fellowship Program. MLGM acknowledges financial support from the grant CEX2021-001131-S funded by MCIU/AEI/ 10.13039/501100011033, from the coordination of the participation in SKA-SPAIN, funded by the Ministry of Science, Innovation and Universities (MCIU), as well as NSERC via the Discovery grant program and the Canada Research Chair program. JHL is supported through the NSERC Discovery grant program, the Discovery Accelerator Supplements program and the Canada Research Chair program.  This manuscript is based on data obtained with the International LOFAR Telescope (ILT). LOFAR \citep{vanhaarlem+13} is the Low Frequency Array designed and constructed by ASTRON. It has observing, data processing, and data storage facilities in several countries, which are owned by various parties (each with their own funding sources), and which are collectively operated by the ILT foundation under a joint scientific policy. The ILT resources have benefited from the following recent major funding sources: CNRS-INSU, Observatoire de Paris and Universit\'e d'Orl\'eans, France; BMBF, MIWF-NRW, MPG, Germany; Science Foundation Ireland (SFI), Department of Business, Enterprise and Innovation (DBEI), Ireland; NWO, The Netherlands; The Science and Technology Facilities Council, UK; Ministry of Science and Higher Education, Poland; The Istituto Nazionale di Astrofisica (INAF), Italy. This research made use of the Dutch national e-infrastructure with support of the SURF Cooperative (e-infra 180169) and the LOFAR e-infra group. The J{\"u}lich LOFAR Long Term Archive and the German LOFAR network are both coordinated and operated by the J{\"u}lich Supercomputing Centre (JSC), and computing resources on the supercomputer JUWELS at JSC were provided by the Gauss Centre for Supercomputing e.V. (grant CHTB00) through the John von Neumann Institute for Computing (NIC). This research made use of the University of Hertfordshire high-performance computing facility and the LOFAR-UK computing facility located at the University of Hertfordshire and supported by STFC [ST/P000096/1], and of the Italian LOFAR IT computing infrastructure supported and operated by INAF, and by the Physics Department of Turin university (under an agreement with Consorzio Interuniversitario per la Fisica Spaziale) at the C3S Supercomputing Centre, Italy.
The National Radio Astronomy Observatory is a facility of the National Science Foundation operated under cooperative agreement by Associated Universities, Inc. The scientific results reported in this manuscript are based in part on data obtained from the Chandra Data Archive.

This paper is based in part on data from the Hyper Suprime-Cam Legacy Archive (HSCLA), which is operated by the Subaru Telescope. The original data in HSCLA was collected at the Subaru Telescope and retrieved from the HSC data archive system, which is operated by the Subaru Telescope and Astronomy Data Center at National Astronomical Observatory of Japan. The Subaru Telescope is honored and grateful for the opportunity of observing the Universe from Maunakea, which has the cultural, historical and natural significance in Hawaii. This paper makes use of software developed for the Vera C. Rubin Observatory. We thank the observatory for making their code available as free software at  \url{http://dm.lsst.org}. This work has made use of the Early Release Observations (ERO) data from the {\it Euclid} mission of the European Space Agency (ESA), 2024,
\url{https://doi.org/10.57780/esa-qmocze3}.
\end{acknowledgements}

%
   \bibliographystyle{aa} 
   \bibliography{refs.bib} 
%

\begin{appendix} 
\section{Spectral index uncertainty maps}
\begin{figure}[h!]
\centering
\includegraphics[width=0.49\textwidth]{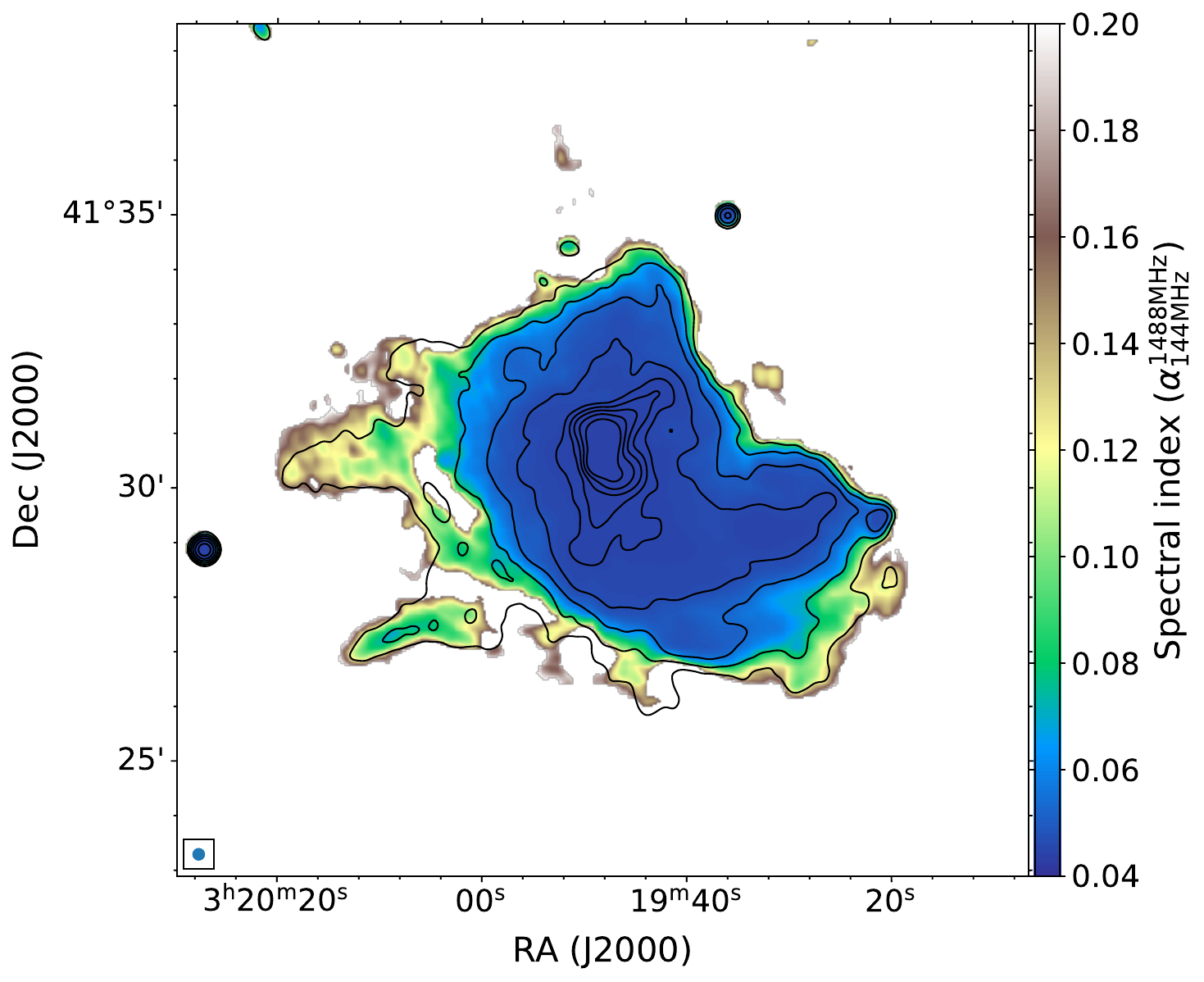}
\includegraphics[width=0.49\textwidth]{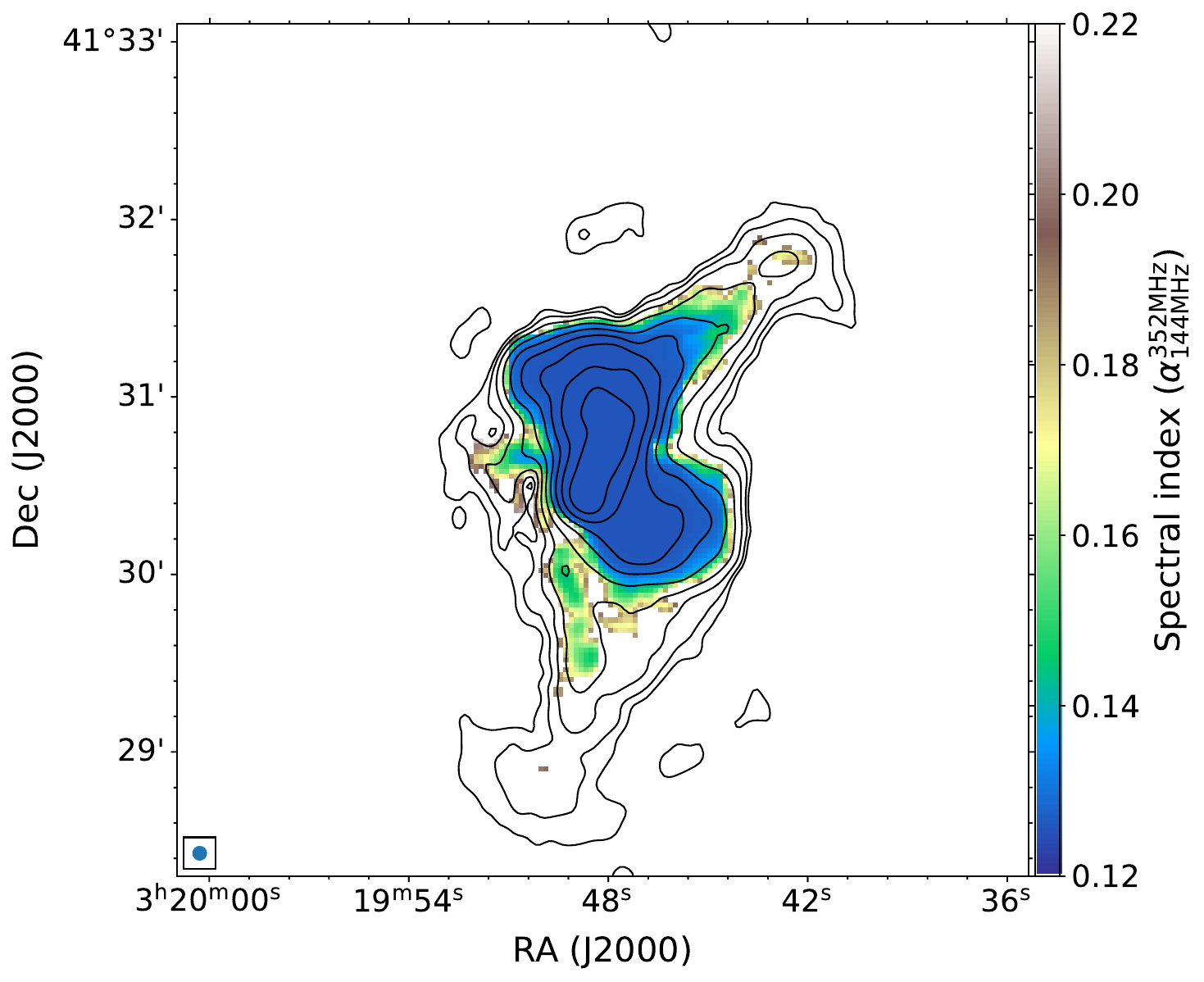}
\caption{Spectral index uncertainty maps corresponding to  Fig.~\ref{fig:spix}. The spectral index uncertainty includes the r.m.s map noise as well as an absolute flux scale uncertainty.}
\label{fig:spixerr}
\end{figure}

The spectral index uncertainty maps corresponding to Fig.~\ref{fig:spix} are shown in Fig.~\ref{fig:spixerr}.

\FloatBarrier

\section{Gaussian gradient magnitude filtered images}
\label{sec:appendixGGM}

\begin{figure}[h!]
\centering
\includegraphics[width=0.49\textwidth]{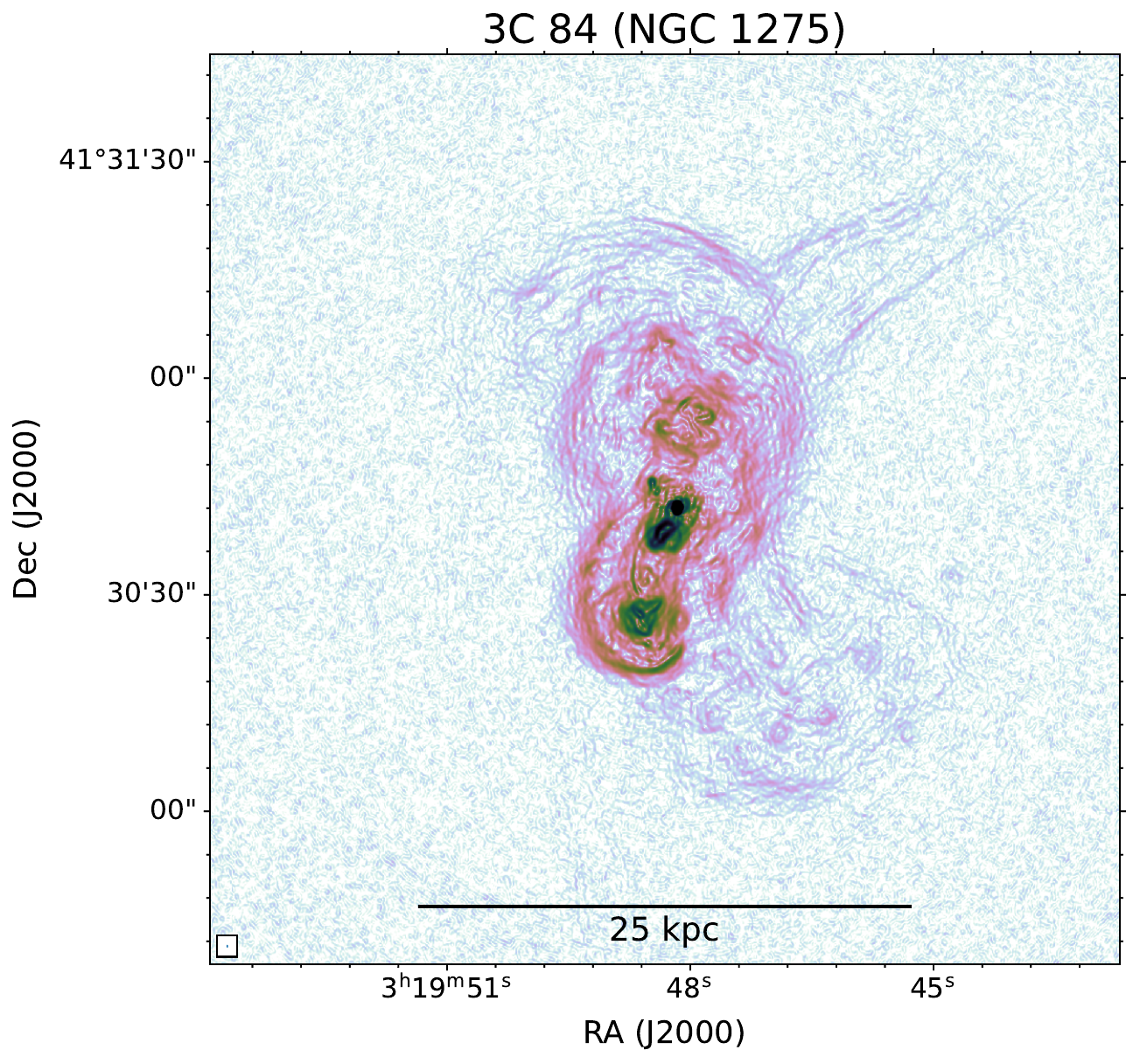}
\includegraphics[width=0.49\textwidth]{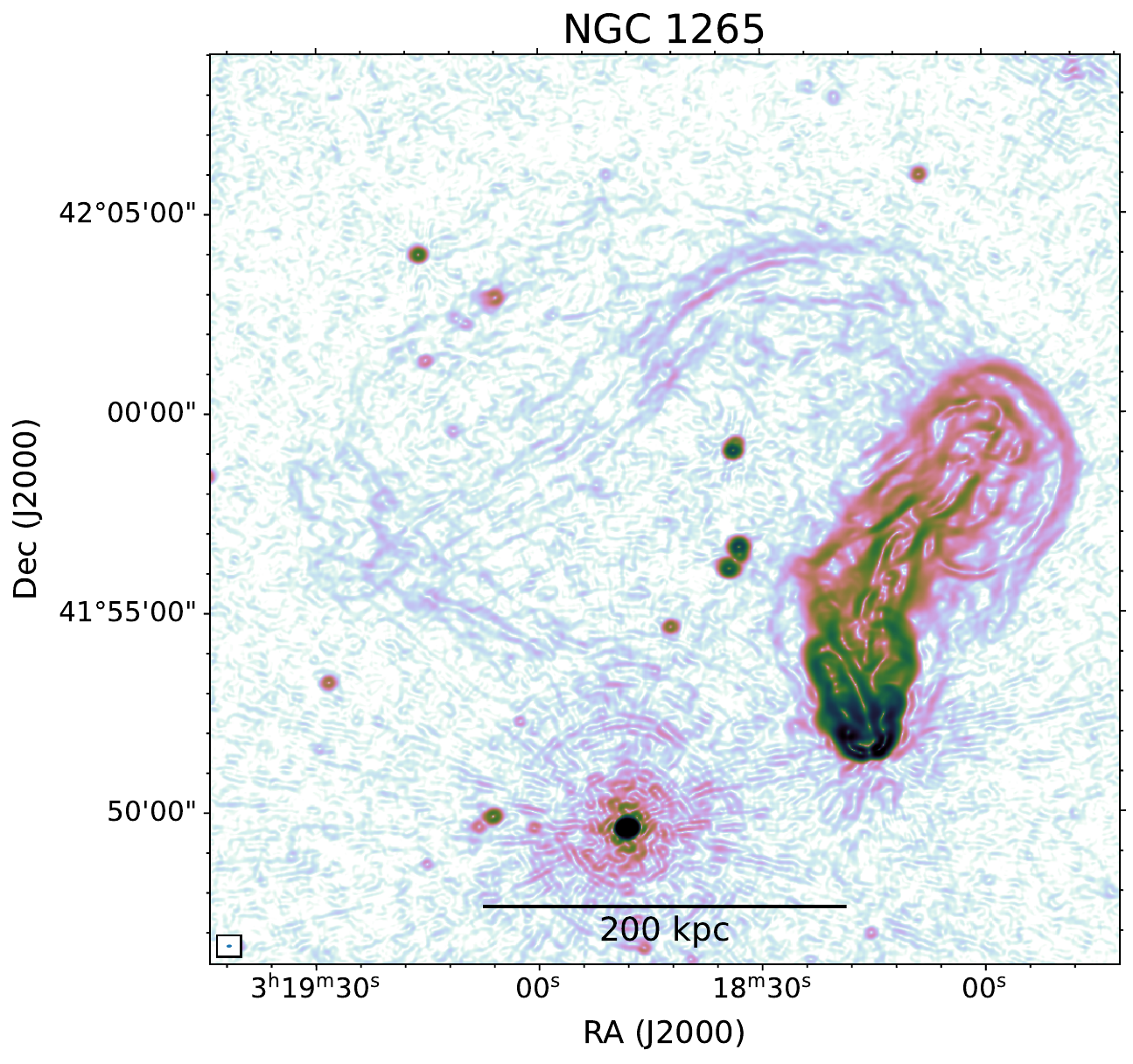}
\caption{GGM filtered images for 3C\,84 (0.3\arcsec{} resolution; top panel) and NGC\,1265 (7\arcsec{} resolution; bottom panel) using $\sigma$ values of 0.1875\arcsec{} and 3.75\arcsec{}, respectively.}
\label{fig:GGMappendix}
\end{figure}

We show the Gaussian gradient magnitude (GGM) filtered images of 3C\,84 (NGC\,1275) at 0.3\arcsec{} resolution in Fig.~\ref{fig:GGMappendix} (left panel). The 7\arcsec{} resolution GGM image for NGC\,1265 is shown in Fig.~\ref{fig:GGMappendix} (right) panel.
\FloatBarrier

\section{Compact source-subtracted images}
\begin{figure*}[t!]
\centering
\includegraphics[width=0.49\textwidth]{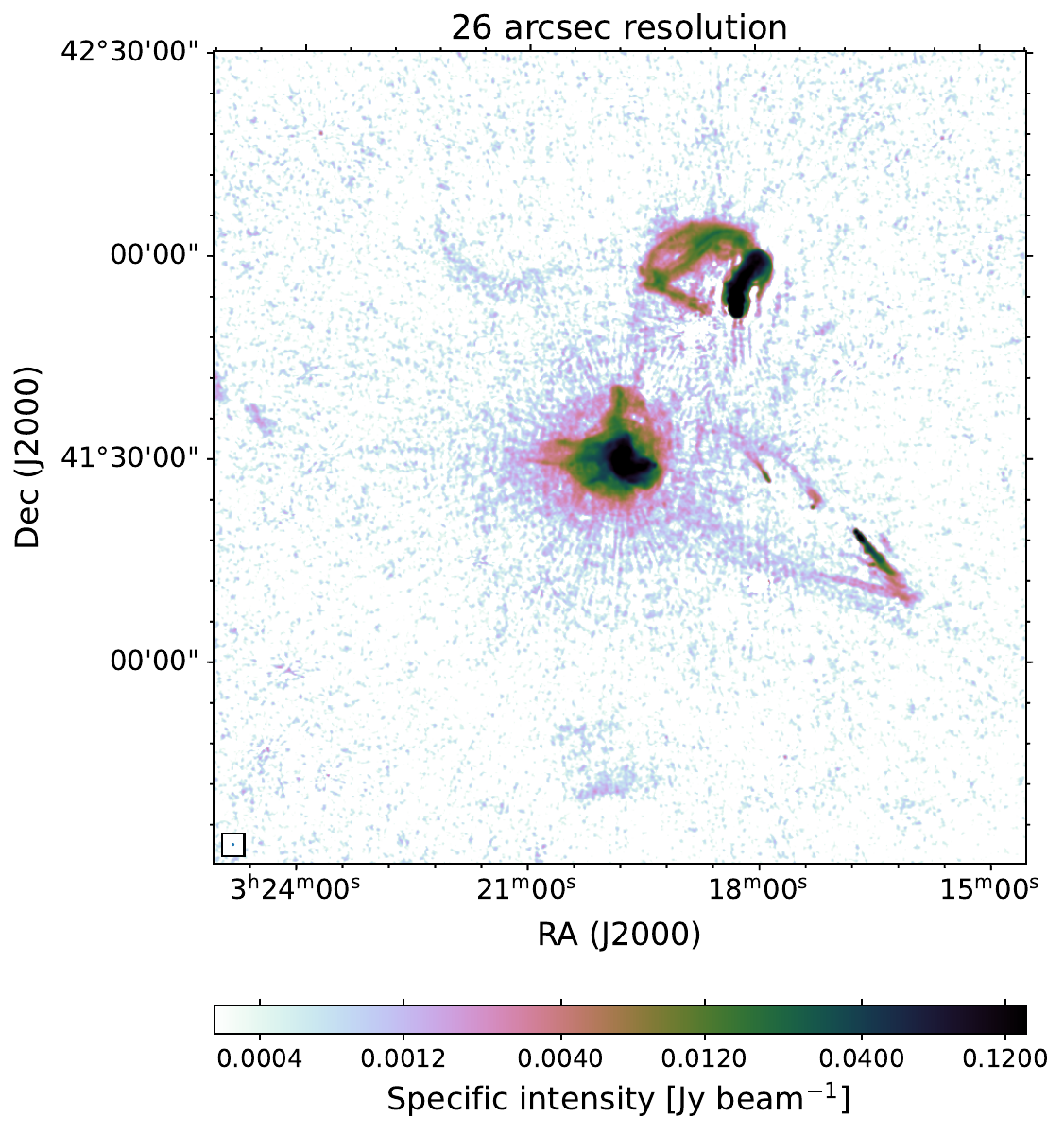}
\includegraphics[width=0.49\textwidth]{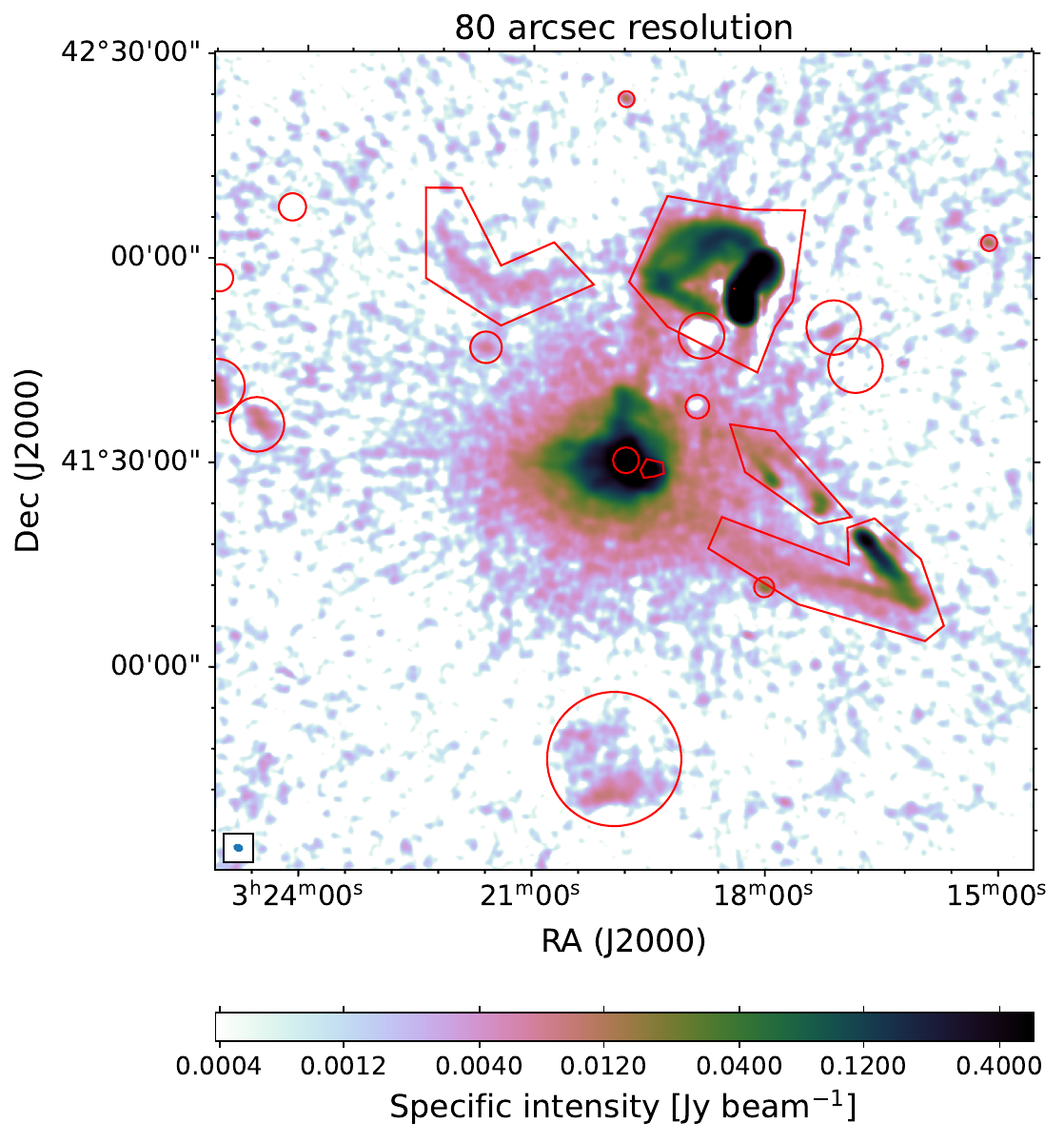}
\caption{Compact source-subtracted images. The images were made employing uv-tapers and have a resolution of 26\arcsec$\times$25\arcsec{} (left) and 90\arcsec$\times$70\arcsec{} (right). The red areas show the regions excluded for determining the radial surface brightness shown in Fig.~\ref{fig:radprofile}. 
}
\label{fig:HBAlowressub}
\end{figure*}

We show compact source--subtracted  26\arcsec{} and 80\arcsec{} resolution images of the Perseus cluster in Fig.~\ref{fig:HBAlowressub}.

\FloatBarrier

\section{Background cluster}
\label{sec:backgroundcluster}
\begin{figure}[h!]
\centering
\includegraphics[width=0.5\textwidth]{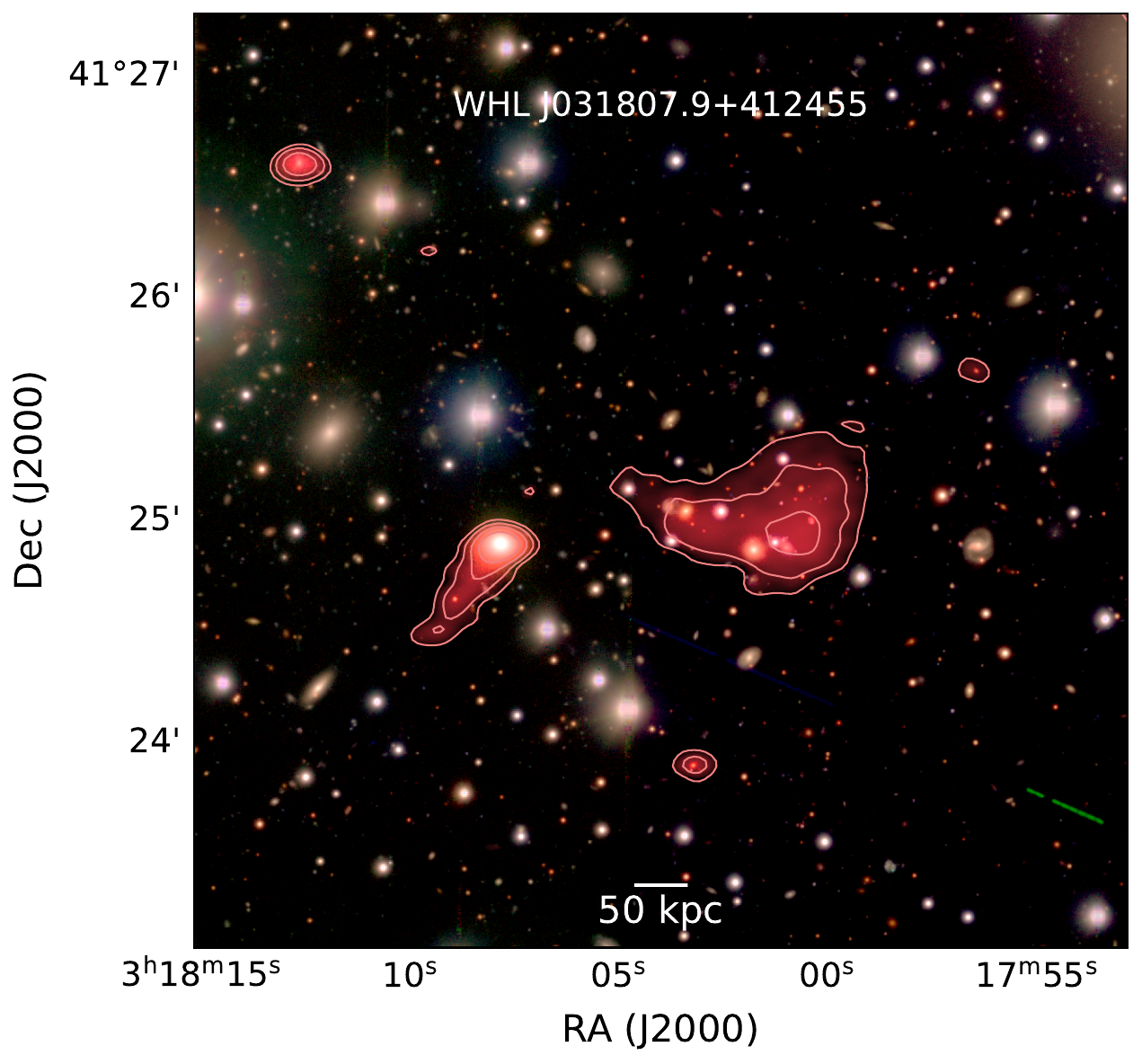}
\caption{Subaru archival $g$, $r$, and~$i$ band color image for the cluster \object{WHL\,J031807.9+412455} situated behind ($z_{\rm{phot}}=0.2125$)  the Perseus cluster  obtained from the Hyper Suprime-Cam (HSC) Legacy
Archive \citep{2021PASJ...73..735T}. The radio emission is overlaid in red from the 7\arcsec{} resolution LOFAR image. Red contours are drawn at levels of $[1,2,4,8] \times 4\sigma_{\rm{rms}}$ and are from the same image.}
\label{fig:backgroundcluster}
\end{figure}

In our images we identify a tailed radio source that is associated with a likely member galaxy \citep[$z_{\rm{phot}}=0.216\pm0.011$][]{2020ApJS..249....3A} from the background galaxy cluster WHL\,J031807.9+412455 \citep[$z_{\rm{phot}}=0.2125$;][]{2012ApJS..199...34W}, see Fig.~\ref{fig:backgroundcluster}. Near the vicinity of this tailed AGN, about an arcminute to the west, a small patch of diffuse emission is detected. This source could be associated with another candidate member galaxy SDSS\,J031801.87+412453.6 \citep[$z_{\rm{phot}}=0.2042\pm0.022$][]{2020ApJS..249....3A}.  \cite{2019ApJS..245...10W} also identify both galaxies as probable background objects to the Perseus cluster. 

\FloatBarrier

\end{appendix}

\end{document}